\documentclass[10pt,aps,prb,twocolumn,floatfix]{revtex4-2}

\usepackage{placeins}
\usepackage{tabularx}
\usepackage{booktabs}

\usepackage{tikz}
\usepackage{pgf}
\usetikzlibrary{decorations}
\usetikzlibrary{decorations.pathreplacing}
\usetikzlibrary{decorations.pathmorphing}

\usepackage{graphicx}
\usepackage{dcolumn}
\usepackage{bm}

\usepackage{qcircuit}
\usepackage{multirow}
\usepackage[caption=false]{subfig}

\usepackage{xcolor}
\usepackage[dvipsnames]{xcolor}  
\usepackage{braket}
\usepackage{array}
\usepackage{amsmath}
\usepackage{amssymb}
\usepackage{amsthm}
\newtheorem{lemma}{Lemma}[section]
\newtheorem{theorem}{Theorem}[section]

\usepackage{threeparttable}
\usepackage[colorlinks=true, allcolors=blue]{hyperref}
\definecolor{bluegreen}{RGB}{0, 139, 150}
\definecolor{dgreen}{RGB}{0, 139, 50}
\begin{document}

\preprint{APS/123-QED}

\title{Max-Cut graph–driven quantum circuit design for geometrically frustrated planar spin systems with spin-glass–like energy landscapes}

\author{Seyed Ehsan Ghasempouri}
\email{ehsan.ghasempouri@unb.ca}
\affiliation{Department of Chemistry, University of New Brunswick, Fredericton, New Brunswick E3B 5A3, Canada}

\author{Gerhard W. Dueck}
\email{gdueck@unb.ca}
\affiliation{Faculty of Computer Science, University of New Brunswick, Fredericton, New Brunswick E3B 5A3, Canada}

\author{Stijn De Baerdemacker}
\email{stijn.debaerdemacker@unb.ca}
\affiliation{Department of Chemistry, University of New Brunswick, Fredericton, New Brunswick E3B 5A3, Canada}
\affiliation{Department of Mathematics and Statistics, University of New Brunswick, Fredericton, New Brunswick E3B 5A3, Canada}

\date{\today}

\begin{abstract}
Finding the ground state of geometrically frustrated spin systems is a challenging problem with broad implications. Many hard optimization problems, including NP-complete problems, can be mapped, for instance, to frustrated Ising models, where competing interactions produce rugged, spin-glass–like energy landscapes. The difficulty is particularly pronounced in the weak-field regime, where geometrical frustration dominates, and the spectral gap becomes exponentially small, making it hard to identify the true ground state. In this work, we consider planar frustrated lattices constructed from the triangular motif, the minimal unit of geometrical frustration. We present a graph-based approach that allows for accurate state initialization of a frustrated triangular spin lattice with up to 20 sites while avoiding barren plateaus. To optimize circuit efficiency and trainability, we employ a clustering strategy that organizes qubits into distinct groups based on the maximum cut technique, which divides the lattice into two maximally disconnected subsets. We provide evidence that this Max-Cut-based lattice division offers a robust framework for optimizing circuit design and effectively modeling frustrated systems at polynomial cost.
All simulations are performed within the variational quantum eigensolver (VQE) formalism, the current paradigm for noisy intermediate-scale quantum (NISQ) devices, but can be extended beyond. Our results underscore the potential of hybrid quantum–classical methods in addressing complex optimization problems.
\end{abstract}

\maketitle


\section{Introduction}
\label{sec:1}
Geometrical frustration is now widely recognized as a central mechanism for
generating rugged, glassy energy landscapes and slow dynamics in interacting
spin systems, even in the absence of quenched disorder, reproducing key
phenomenology traditionally associated with Ising spin glasses
\cite{ramirez1994strongly,zhou2017glassy,moessner1998spinliquid,
chamon2005quantumglass,seidler1996anomalous,mitsumoto2020spinorbital,
ali2024quench,bramwell2001spinice,hen2016driver,mila2015frustrated}.

Determining the ground state of an Ising spin glass \cite{edwards1975theory}, a problem within the class of combinatorial optimization, becomes exponentially challenging for classical computers as the system size increases, similar to other many-body problems. Quantum computers, first proposed in the early 1980s \cite{Benioff,Manin,Feynman}, are anticipated to surpass these limitations by leveraging quantum mechanical phenomena, including superposition, interference, and entanglement. This advancement in quantum computing offers significant potential for solving specific computational challenges. Two of the most prominent algorithms in this field are the quantum phase estimation (QPE) \cite{Kitaev,Kitaev_Alexei,Cleve,Nielsen,Abrams} and the variational quantum eigensolver (VQE) \cite{Peruzzo,McClean}. The QPE algorithm typically requires a high circuit depth to accurately determine the ground state energy, which necessitates further development in error-corrected hardware. Consequently, the VQE, a quantum-classical hybrid approach with shallow circuit depth, is a viable alternative for noisy intermediate-scale quantum (NISQ) devices~\cite{Preskill}. However, this advantage comes at the cost of potentially encountering an NP-hard optimization problem during the training process~\cite{bittel2021training}. The objective of the VQE is to find a parameterized ansatz that can determine the energy of the target state or provide a reasonable quantum state estimation for the QPE. The ansatz is prepared by designing a quantum circuit with a variational form, consisting of layers that specify rotation patterns and CNOT gates. One of the most challenging aspects of VQE is the design of an efficient variational quantum circuit with shallow depths, which is crucial in noisy environments and demands a minimal number of parameters to avoid issues related to barren plateaus and entrapment in local minima. 

The Ising problem has a tight connection with optimization problems. The Quantum Approximate Optimization Algorithm (QAOA), introduced by Farhi, Goldstone, and Gutmann in 2014 \cite{farhi2014quantum}, is an optimization algorithm based on the VQE method designed to address combinatorial problems. The QAOA has been widely studied to solve quadratic unconstrained binary optimization (QUBO) problems \cite{BLEKOS20241,moussa2022unsupervised} such as   Maximum Cut (Max-Cut) \cite{farhi2014quantum,basso2021quantum}, Maximum Independent Set (MIS) \cite{choi2019tutorial},  Binary Paint Shop Problem (BPSP) \cite{streif2021beating}, Binary Linear Least Squares (BLLS) \cite{borle2021quantum}, etc. The QAOA method has many scientific and industrial applications such as portfolio optimization \cite{hodson2019portfolio},  tail assignment \cite{vikstaal2020applying},  protein folding \cite{mustafa2022variational}, etc. In several situations, such as QUBO, the QAOA can be mapped to the problem of finding the ground state of an Ising system.

Exploring the ground-state properties of Ising spin glasses stands as a key difficulty in both solid-state physics and statistical mechanics and has been extensively researched for many decades \cite{binder1986spin,mezard1987spin,mezard2009information,stein2013spin}. The term spin glass denotes a state of disordered magnetism where the spins of magnetic atoms or molecules interact in such a way that it gives rise to frustrated magnetic phenomena \cite{mydosh1993spin}. The behavior of frustrated magnetic metals is observed in alloys such as CuMn and AuFe, where manganese (Mn) or iron (Fe) is introduced as impurities into copper (Cu) or gold (Au) \cite{cannella1972magnetic,mulder1981susceptibility}. The Ising model is widely used due to its simplicity and effectiveness in capturing the essential features of real spin glasses, as it accounts for the interaction energy between impurities~\cite{de1995exact,barahona1982computational}. The presence of unsatisfied edges in the lattice can lead to situations where multiple geometric arrangements of atoms remain stable. For a spin glass with $n$ spins, there are $2^n$  possible configurations, making the task of finding the ground state a combinatorial problem. It has been shown that for planar graphs with nearest-neighbor interactions, both with and without periodic boundary conditions, and in the absence of an external magnetic field, the ground state can be found in polynomial time \cite{edmonds1965minimum,bieche1980ground,barahona1982morphology,d1984random,hadlock1975finding,orlova1972finding} and is comparable to solving the Max-Cut problem. However, Barahona proved that computing the ground state energy for two-dimensional (2D) models with a magnetic field is NP-hard, even without periodic boundary conditions, and no polynomial algorithm exists for this problem \cite{barahona1982computational}.

Over the past few decades, various methods, such as Monte Carlo simulations \cite{de1996exact} and evolutionary algorithms \cite{gropengiesser1995ground,gropengiesser1995superlinear}, including genetic algorithms \cite{sutton1994genetic}, have been developed to determine the ground state energy of Ising spin glasses. However, these approaches generally provide only approximations rather than the exact ground state, with a significant limitation that there is no guarantee of finding the true ground state \cite{barahona1994ground,de1995exact}. Additionally, another major drawback of these heuristic methods is their performance in the presence of degeneracies. Two states with almost identical energy levels can still be qualitatively different. Therefore, even though these algorithms can give energies close to the ground state, they may not be close to the true ground state. As a result, these methods are inadequate for analyzing real ground states, such as those found in protein folding applications, as mentioned in reference \cite{liang1992application}. However, finding the exact ground state of a 2D Ising model with a transverse field is itself an NP-hard problem, and thus no polynomial-time algorithm—classical or quantum—can be expected to reliably produce exact solutions for general instances.

An alternative approach to finding the true ground state involves adiabatic methods, such as simulated annealing (SA) \cite{kirkpatrick1983optimization}, quantum annealing (QA) \cite{johnson2011quantum}, and the quantum adiabatic algorithm (QAA) \cite{farhi2000quantum}. The SA method is an optimization technique that seeks the ground state by simulating a system gradually cooling into a low-energy state. The downside of this method is that the cooling process must be prolonged; otherwise, the system may get trapped in local minima rather than reaching the global minimum. The QA relies on quantum fluctuations and aims to find optimal solutions by gradually reducing an applied transverse magnetic field to the spin system at the lowest possible temperature. Experiments \cite{brooke1999quantum} and simulations \cite{kadowaki1998quantum,santoro2002theory} have demonstrated that QA can achieve equilibrium in a spin glass more rapidly than SA, which operates through thermal annealing. However, removing the magnetic field reduces quantum fluctuations and forces the system to pass through the critical point. A complete understanding of the transition between the disordered phase, which occurs at large fields, and the spin glass phase, which emerges at smaller fields, is still lacking and remains an active area of research. A recent study \cite{bernaschi2024quantum} utilizing extreme-scale simulations, primarily based on Monte Carlo methods, demonstrated that it is possible to reach the quantum spin-glass phase with an annealing time that scales polynomial in the number of qubits, provided that parity-changing excitations are avoided. One possible limitation of this optimistic outlook is that simply crossing the critical point differs from navigating through the entire bulk glass phase to reach the final state at zero transverse-field, which may still be challenging \cite{bode2024adiabatic}. The QAA is closely related to the QA but is specifically a Hamiltonian-based quantum algorithm where the Hamiltonian undergoes slow evolution, enabling the system to remain in its ground state. Determining the complexity of the QAA is crucial, particularly when the minimum of the first excitation gap ($\Delta{E}_{min}$) is small, since the total evolution time ($\tau$) is inversely related to $\Delta{E}_{min}$. If $\Delta{E}_{min}$ becomes exponentially small at any point during the evolution, the simulation time can grow exponentially, rendering the QAA inefficient \cite{farhi2012performance}. The same issue arises in QA as well, since it also requires adiabatically transitioning from the critical point to a vanishing magnetic field. The complexity of this task becomes even more problematic in higher dimensions \cite{hen2011exponential,knysh2016zero,tanaka2017quantum2}. Related annealing-inspired approaches, such as Snapshot-QAOA
\cite{tate2024snapshot}, rely on gap-preserving interpolations and may therefore encounter similar challenges in frustrated regimes where the many-body excitation gap becomes very small. 

Given the limitations of adiabatic approaches, exploring alternative methods is important. For instance, an algebraic approach has been proposed for designing a quantum circuit that efficiently diagonalizes the Ising Hamiltonian and provides access to all eigenstates by directly preparing the computational basis states. This approach has been applied to a 4-qubit system and holds the potential for extension to larger qubit numbers; however, only for the one-dimensional Ising model \cite{verstraete2009quantum,cervera2018exact}. Recently, a promising study~\cite{kim2023evidence} demonstrated the effectiveness of noise-mitigation techniques in quantum simulations of a 2D transverse-field Ising model implemented on a 127-qubit heavy-hexagon lattice. The results surpassed those of several classical tensor network methods applied to the same problem, even when the classical approaches utilized substantial computational resources. In addition, in a recent advance, King et al.~\cite{doi:10.1126/science.ado6285} used a large-scale, programmable quantum annealer to analyze entanglement growth in spin glasses, demonstrating area-law scaling in quench dynamics across two-, three-, and infinite-dimensional topologies. They showed that state-of-the-art classical methods based on tensor and neural networks could not match the quantum annealer’s accuracy within a reasonable timeframe.

From a complexity-theoretic perspective, the Ising spin-glass ground-state task naturally appears in two forms: an optimization version (compute the minimum energy) and a decision version (given a threshold $E_0$, decide whether the ground-state energy satisfies $E_{\min}\le E_0$). The former is NP-hard~\cite{barahona1982computational}, whereas the latter is NP-complete \cite{Lucas_2014}. Consequently, the decision form is polynomial-time interreducible with any NP-complete problem, formalizing the close connection between Ising models and a broad class of hard combinatorial decision problems. This connection helps explain why advances in algorithms for low-energy Ising configurations can have wide impact: many challenging optimization problems---including the traveling salesman problem and protein folding---admit Ising (or closely related) formulations \cite{Lucas_2014,FORTUNATO201075,goldstein1992optimal}.
While it is widely conjectured that quantum computers will not solve NP-complete problems in polynomial time, they may still yield practically relevant speedups---for instance, by reducing the exponent in the best-known exponential scaling for certain problem families, as suggested by recent QAOA scaling studies for random 8-SAT \cite{BoulebnaneMontanaro2024PRXQuantum}.
The broader problem-solving potential that comes with determining low-energy configurations of Ising models motivates researchers from diverse disciplines to tackle this challenge. For instance, the Hopfield model \cite{hopfield1982neural}, along with other neural network models, uncovered significant links between neural networks and Ising spin glasses across general networks~\cite{little1974existence,amit1985spin,sompolinsky1988statistical}. A range of deep neural network architectures, including generative models~\cite{wu2019solving,hibat2021variational,mcnaughton2020boosting,gabrie2022adaptive,
pan2021solving,van2016pixel,ma2024message} and reinforcement learning~\cite{mills2020finding,fan2023searching}, have been applied to solve Ising models. Designing neural networks to solve the Ising spin glass model on larger and more complex graphs can not only open the door for the development of powerful optimization tools but also illuminate the limitations and boundaries of deep-learning-assisted algorithms \cite{ma2024message}.

In light of the computational hardness of frustrated Ising models in the
presence of a transverse field, our goal is not to provide an exact
polynomial-time algorithm for arbitrary systems. Instead, we position our
contribution as a structure-guided heuristic ansatz for the VQE that explicitly leverages the geometry and symmetries of the underlying lattice. By combining (i) Max-Cut-based bipartitions that separate the lattice into minimally frustrated groups, (ii) a symmetry-preserving two-qubit cluster (2-qc) building block, and (iii) a circuit layout aligned with the global spin-flip symmetry of the Hamiltonian, we construct variational circuits that remain confined to the physically relevant subspace and substantially mitigate the risk of barren plateaus. As demonstrated throughout this work, this structure-exploiting design leads to markedly improved trainability, high ground-state fidelity, and polynomial circuit scaling, even
in regimes where standard VQE, QAOA~\cite{farhi2014quantum}, or
ADAPT-VQE~\cite{grimsley2019adaptive} struggle.

In earlier works, maximum-weight cuts were employed to compute ground-state energies of two-dimensional Ising spin-glass models with random Gaussian couplings on square lattices using classical hardware \cite{barahona1994ground,de1995exact}. Subsequent studies considered the more challenging \(\pm J\) case, where the interaction magnitude is fixed but the sign of each bond is random \cite{de1996exact}. Using branch-and-cut techniques, instances as large as a \(100 \times 100\) grid can be treated; however, the runtime is difficult to predict, and for larger lattices, obtaining exact solutions with guaranteed polynomial scaling remains challenging. In the present work, rather than introducing disorder through random couplings, we focus on geometrically frustrated lattices with fixed antiferromagnetic interactions \(J_{ij}<0\) and use Max-Cut as a structural guideline for designing efficient quantum circuits. In the weak-field regime, this geometrical frustration gives rise to complex low-energy landscapes that share qualitative features with those reported in disordered spin-glass systems.

In this study, we apply the proposed ansatz to frustrated triangular spin lattices with system sizes up to 20 sites and employ VQE to compute ground state energies. We show that the Max-Cut realization of the lattice provides a natural and effective guideline for partitioning the system into two groups and establishing the corresponding inter-cluster couplings.

The remainder of the paper is organized as follows. We first introduce the concept of geometrical frustration and develop an exact theoretical framework for the triangular transverse-field Ising model. As the minimal frustrated motif, the triangle serves as a fundamental building block of the triangular lattice and is described using a cluster-circuit construction based on graph partitioning. We then extend this framework to study frustrated systems with 5 and 7 sites, as well as a 10-site triangular lattice. Finally, we explore the applicability of the proposed approach to predicting the magnetic behavior of larger frustrated lattices, with system sizes up to 20 sites.

\section{Geometrical frustration}
\label{sec:geometrical_frustration}

\begin{figure}[htbp]
    \centering
    \includegraphics[width=0.3\textwidth]{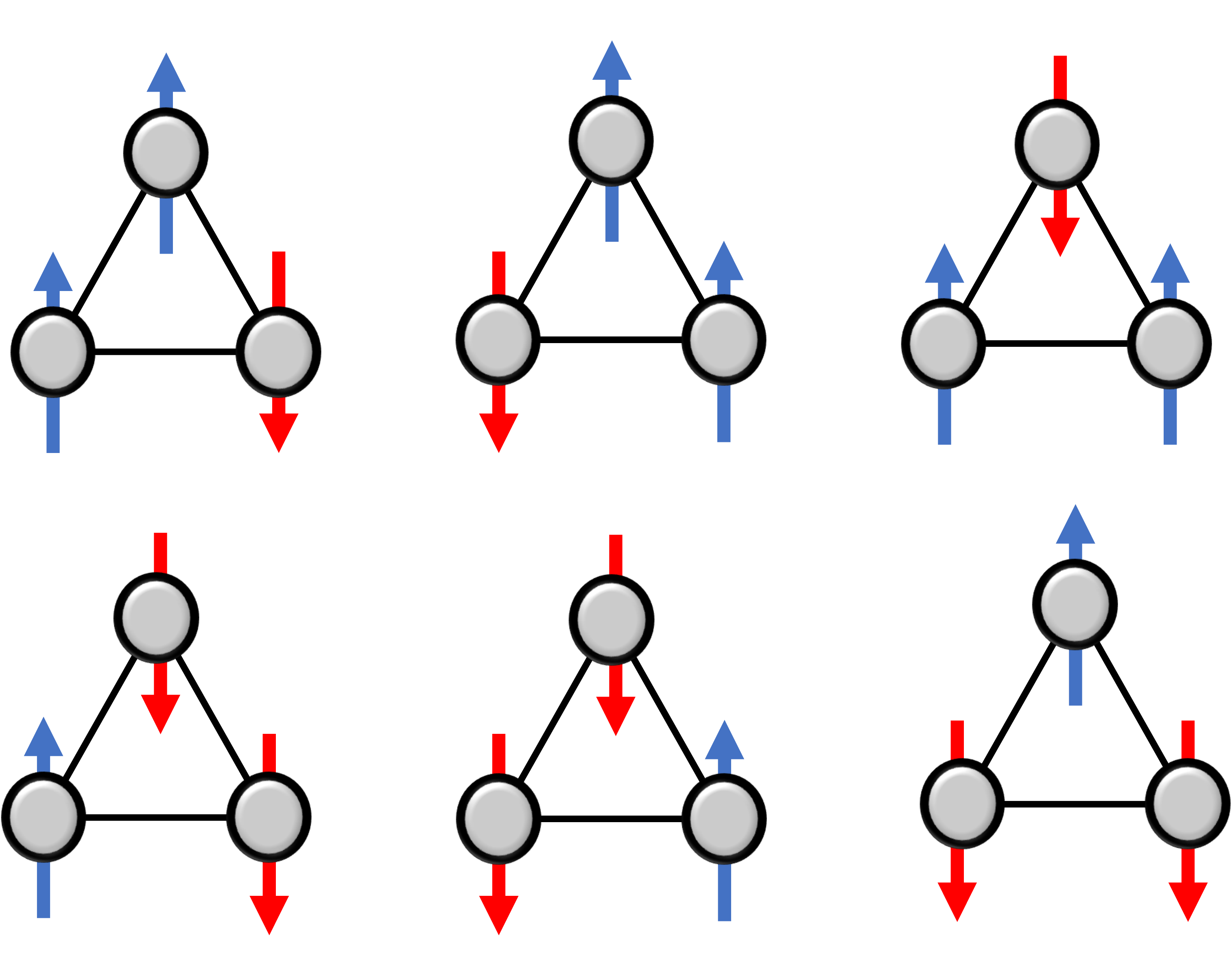}
    \caption{Six degenerate AF Ising ground states on a triangle.}
    \label{fig.2} 
\end{figure}

Geometrical frustration arises when the interaction pattern in a spin system prevents all pairwise antiferromagnetic (AF) couplings from being satisfied simultaneously. The triangular motif is the simplest example: for AF Ising interactions, at most two spins can align antiparallel, leaving one bond necessarily unsatisfied. This constraint produces six degenerate classical ground-state configurations, shown in Fig.~\ref{fig.2}.

This multiplicity of low-energy configurations is a defining feature of
geometrical frustration and becomes increasingly pronounced in larger frustrated
lattices.

\subsection{Transverse-field Ising model}
\label{subsec:TFIM}

To study quantum effects in frustrated systems, we consider the
transverse-field Ising model (TFIM),
\begin{equation}
    H = -\sum_{\langle i,j\rangle} J_{ij} Z_i Z_j - h \sum_i X_i ,
    \label{eq:TFIM}
\end{equation}
where \( J_{ij}<0 \) encodes the AF Ising interactions and \(h\) represents the transverse magnetic field. Introducing the dimensionless parameters $a,b$ with $a \in [0,1]$ and $b = 1 - a$, we set $J_{ij} = -b$ and $h = -a$ in Eq.~(\ref{eq:TFIM}). Thus, $a$ controls the transverse field and $b$ controls the antiferromagnetic interaction strength. The transverse field introduces quantum fluctuations by flipping individual spins, thereby mixing the many degenerate classical configurations induced by frustration. The simulated lattice geometries considered in this work are shown in Fig.~\ref{fig:lattices}.
\begin{figure}[htbp]
    \centering

    \begin{minipage}{0.15\textwidth}
        \centering
    {\color{Orchid}
    \setlength{\fboxrule}{1.2pt}
    \setlength{\fboxsep}{6pt}
        \fbox{%
        \begin{minipage}{0.9\textwidth}
{\Large \textcolor{Orchid}{House}}

        \vspace{1.2em}
            \centering
            \begin{tikzpicture}[scale=0.2, every node/.style={circle, draw=Orchid, line width=1pt, minimum size=0.2cm, fill=white}]
                \node (n0) at (0, 0) {};
                \node (n1) at (6,0) {};
                \node (n2) at (0,6) {};
                \node (n3) at (6,6) {};
                \node (n4) at (3,10) {};
                \draw[Orchid, very thick]
                    (n0)--(n1) (n0)--(n2) (n2)--(n3)
                    (n1)--(n3) (n2)--(n4) (n3)--(n4);
            \end{tikzpicture}

        \end{minipage}
        }
        }
    \end{minipage}
    \hfill
    \begin{minipage}{0.25\textwidth}
        \centering
                {\color{YellowOrange}
    \setlength{\fboxrule}{1.2pt}
    \setlength{\fboxsep}{6pt}
        \fbox{%
        \begin{minipage}{0.9\textwidth} 
        {\Large \textcolor{YellowOrange}{Hexagon}}
        
\vspace{1.2em}
            \centering
            \begin{tikzpicture}[scale=0.11, every node/.style={circle, draw=YellowOrange, line width=1pt, minimum size=0.2cm, fill=white}]
                \node (n0) at (-10.5, 0) {};
                \node (n1) at (-7,-7) {};
                \node (n2) at (0,-7) {};
                \node (n3) at (3.5,0) {};
                \node (n4) at (0,7) {};
                \node (n5) at (-7,7) {};
                \node (n6) at (-3.5,0) {};
                \draw[YellowOrange, very thick]
                (n0)--(n1) (n0)--(n5) (n0)--(n6)
                      (n1)--(n2) (n1)--(n6)
                      (n2)--(n3) (n2)--(n6)
                      (n3)--(n4) (n3)--(n6)
                      (n4)--(n5) (n4)--(n6)
                      (n5)--(n6);
            \end{tikzpicture}


            \vspace{1.8em}

            \begin{tikzpicture}[scale=0.8, every node/.style={circle, draw=YellowOrange,line width=1pt,  minimum size=0.2cm, fill=white}]
                \node (0)  at (0,4) {};
                \node (1)  at (1,4) {};
                \node (2)  at (2,4) {};

                \node (3)  at (-0.5,3) {};
                \node (4)  at (0.5,3) {};
                \node (5)  at (1.5,3) {};
                \node (6)  at (2.5,3) {};

                \node (7)  at (-1,2) {};
                \node (8)  at (0,2) {};
                \node (9)  at (1,2) {};
                \node (10) at (2,2) {};
                \node (11) at (3,2) {};

                \node (12) at (-0.5,1) {};
                \node (13) at (0.5,1) {};
                \node (14) at (1.5,1) {};
                \node (15) at (2.5,1) {};

                \node (16) at (0,0) {};
                \node (17) at (1,0) {};
                \node (18) at (2,0) {};

                \draw[YellowOrange, very thick]
                (0)--(1) (0)--(4) (0)--(3)
                      (1)--(2) (1)--(4) (1)--(5)
                      (2)--(5) (2)--(6)
                      (3)--(4) (3)--(7) (3)--(8)
                      (4)--(5) (4)--(8) (4)--(9)
                      (5)--(6) (5)--(9) (5)--(10)
                      (6)--(10) (6)--(11)
                      (7)--(8) (7)--(12)
                      (8)--(12) (8)--(13)
                      (8)--(9)
                      (9)--(10) (9)--(13) (9)--(14)
                      (10)--(11) (10)--(14) (10)--(15)
                      (11)--(15)
                      (12)--(13) (12)--(16)
                      (13)--(14) (13)--(16) (13)--(17)
                      (14)--(15) (14)--(17) (14)--(18)
                      (15)--(18)
                      (16)--(17)
                      (17)--(18);
            \end{tikzpicture}

        \end{minipage}
        }
        }
    \end{minipage}
    \hfill
    \vspace{1.5em}
    \begin{minipage}{0.4\textwidth}
        \centering
                {\color{teal}
    \setlength{\fboxrule}{1.2pt}
    \setlength{\fboxsep}{6pt}
        \fbox{%
        \begin{minipage}{0.9\textwidth}
        {\Large \textcolor{teal}{Ribbon}}
        
        \vspace{1.2em}
            \centering
            \begin{tikzpicture}[scale=0.13, every node/.style={circle, draw=teal,line width=1pt, minimum size=0.2cm, fill=white}]
                \node (n0) at (-10.5, 0)  {};
                \node (n1) at (-7,-7)     {};
                \node (n2) at (0,-7)      {};
                \node (n3) at (7,-7)      {};
                \node (n4) at (10,0)      {};
                \node (n5) at (7,7)       {};
                \node (n6) at (0,7)       {};
                \node (n7) at (-7,7)      {};
                \node (n8) at (-3.5,0)    {};
                \node (n9) at (3.5,0)     {};

                \draw[teal, very thick]
                (n0)--(n1) (n0)--(n7) (n0)--(n8)
                      (n1)--(n2) (n1)--(n8)
                      (n2)--(n3) (n2)--(n8) (n2)--(n9)
                      (n3)--(n4) (n3)--(n9)
                      (n4)--(n5) (n4)--(n9)
                      (n5)--(n6) (n5)--(n9)
                      (n6)--(n7) (n6)--(n8) (n6)--(n9)
                      (n7)--(n8)
                      (n8)--(n9);
            \end{tikzpicture}


            \vspace{1.8em}

            \begin{tikzpicture}[scale=0.13, every node/.style={circle, draw=teal,line width=1pt, minimum size=0.2cm, fill=white}]
                \node (0)  at (-11, 0) {};
                \node (1)  at (-7,-7) {};
                \node (2)  at (0,-7) {};
                \node (3)  at (7,-7) {};
                \node (4)  at (14,-7) {};
                \node (5)  at (21,-7) {};
                \node (6)  at (28,-7) {};
                \node (7)  at (31,0) {};
                \node (8)  at (35,7) {};
                \node (9)  at (28,7) {};
                \node (10) at (21,7) {};
                \node (11) at (14,7) {};
                \node (12) at (7,7) {};
                \node (13) at (0,7) {};
                \node (14) at (-7,7) {};
                \node (15) at (-4,0) {};
                \node (16) at (3,0) {};
                \node (17) at (10,0) {};
                \node (18) at (17,0) {};
                \node (19) at (24,0) {};

                \draw[teal, very thick]
                (0)--(1) (0)--(14) (0)--(15)
                      (1)--(2) (1)--(15)
                      (2)--(3) (2)--(15) (2)--(16)
                      (3)--(4) (3)--(16) (3)--(17)
                      (4)--(5) (4)--(17) (4)--(18)
                      (5)--(6) (5)--(18) (5)--(19)
                      (6)--(7) (6)--(19)
                      (7)--(8) (7)--(9) (7)--(19)
                      (8)--(9)
                      (9)--(10) (9)--(19)
                      (10)--(11) (10)--(18) (10)--(19)
                      (11)--(12) (11)--(17) (11)--(18)
                      (12)--(13) (12)--(16) (12)--(17)
                      (13)--(14) (13)--(15) (13)--(16)
                      (14)--(15)
                      (15)--(16)
                      (16)--(17)
                      (17)--(18)
                      (18)--(19);
            \end{tikzpicture}

        \end{minipage}
        }
        }
    \end{minipage}

    \caption{
Overview of the geometrically frustrated lattices studied in this work. 
Top left: 5-site house geometry. 
Top right: 7-site and 19-site hexagon lattices. 
Bottom: 10-site and 20-site ribbon lattices.
}
    \label{fig:lattices}

\end{figure}

To demonstrate how geometric frustration gives rise to a spin-glass-like energy landscape—characterized by near-degenerate states and multiple local minima—we compare each frustrated geometry with its corresponding unfrustrated chain geometry. This comparison highlights the role of frustration in producing glassy behavior, particularly in the weak-field regime.

\begin{figure}[htbp]
\centering
    \subfloat[]{
    \includegraphics[width=0.47\textwidth]{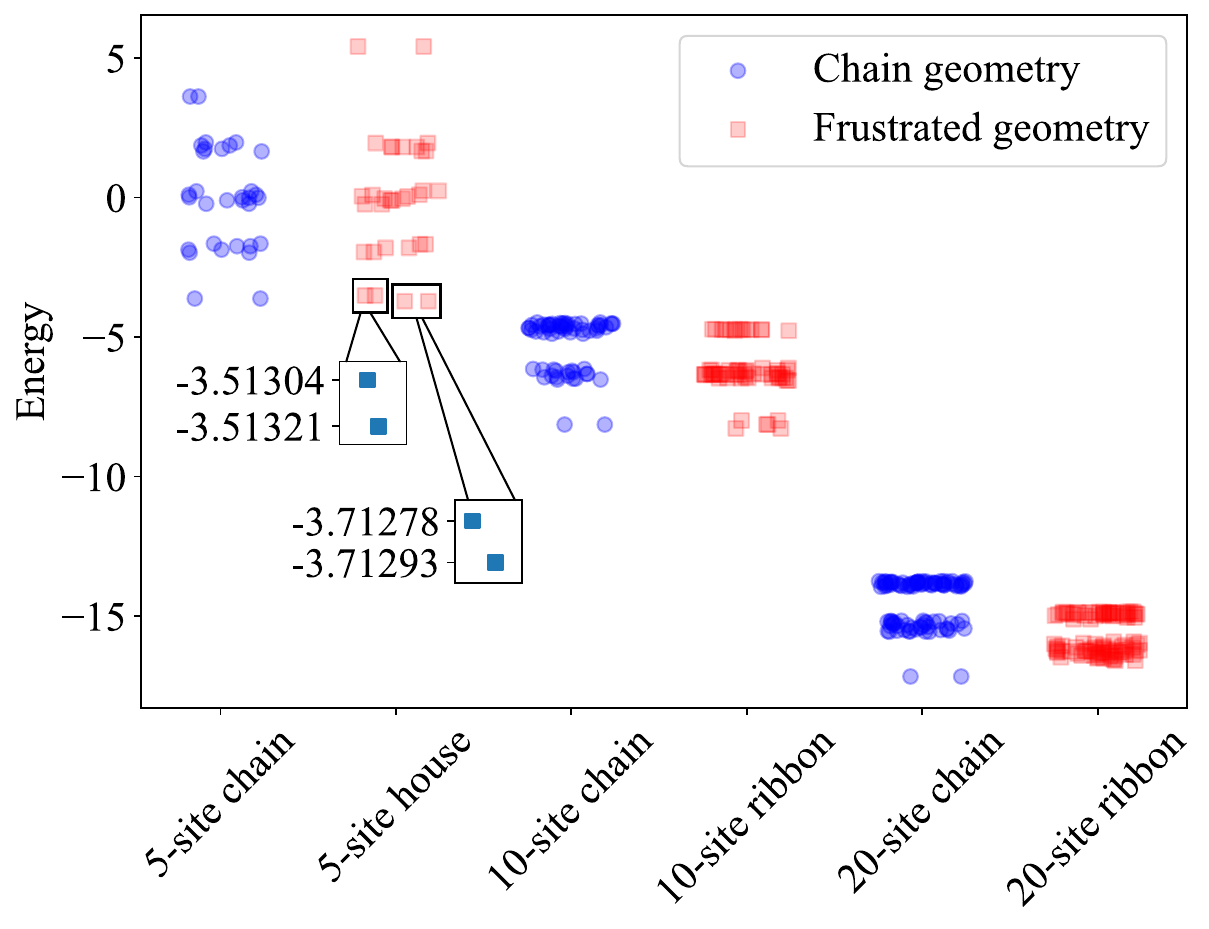}
    \label{fig.gap(a)}
}
\hspace{0.06\textwidth}
    \subfloat[]{
    \includegraphics[width=0.47\textwidth]{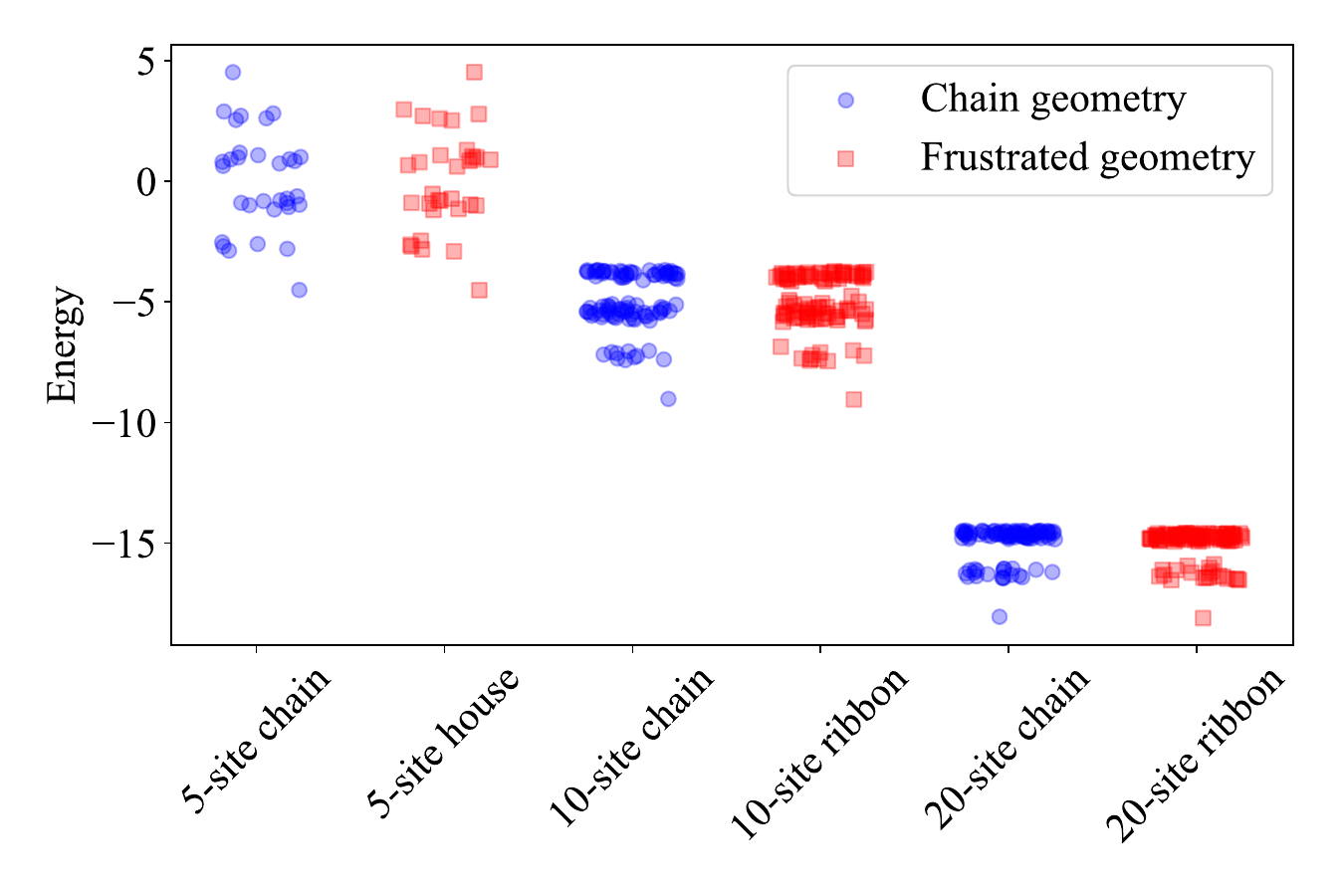}
    \centering
    \label{fig.gap(b)}
    }
\caption{Energy-band diagrams for chain and frustrated geometries with $L=5$ (full spectrum) 
and $L=10,20$ (lowest 100 states).  
(a) Weak-field regime ($a=0.1$).  
(b) Strong-field regime ($a=0.9$).}
\label{fig.gap}
\end{figure}

Fig~\ref{fig.gap} presents the low-energy spectra of both the chain and frustrated geometries for system sizes $L=5,10,20$.  
For $L=5$, the full spectrum ($2^5 = 32$ levels) is shown, whereas for $L=10$ and $20$ we display only the lowest 100 energy levels.

\begin{figure}[htbp]    
   \centering
    \includegraphics[width=0.4\textwidth]{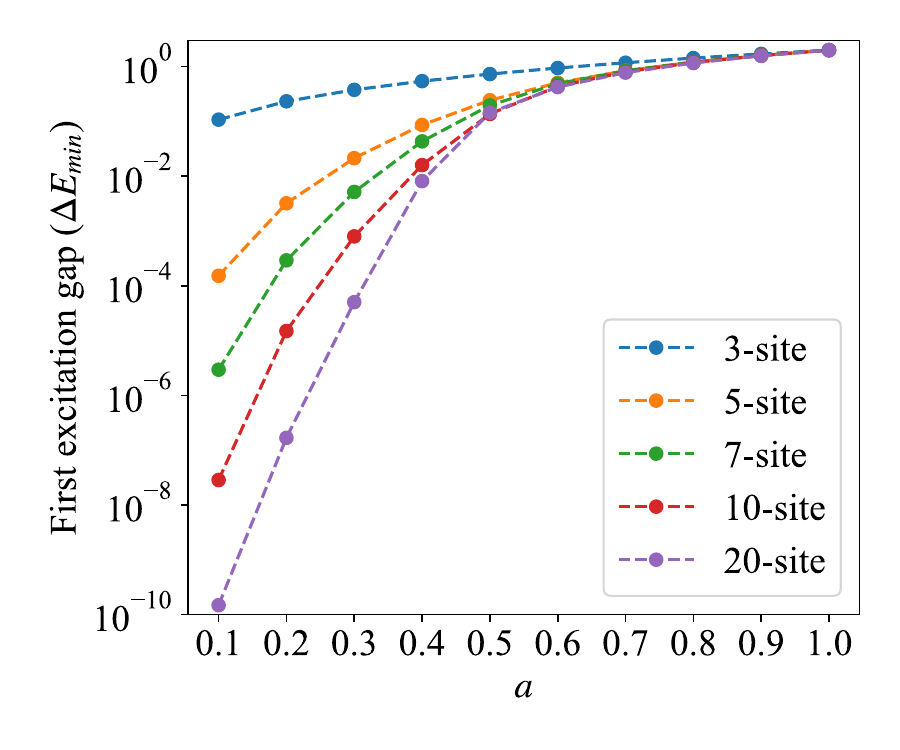}
    \caption{First excitation gap for various lattice sizes and different values of \( a \).}
    \label{fig.17}
\end{figure}

In the \emph{weak-field regime} [Fig.~\ref{fig.gap}(a)], geometric frustration gives rise to a macroscopically near-degenerate ground-state manifold, with eigenstates clustered close to the ground-state energy and separated by a gap that becomes exponentially small with increasing system size. As the system size increases from \( L=5 \) to \( L=20 \), the low-energy spectrum becomes progressively denser and more irregular, indicating the emergence of a rugged energy landscape. The magnified inset of Fig.~\ref{fig.gap}(a), shown for the 5-site case as a representative example, resolves splittings on the order of
\( \mathcal{O}(10^{-4}) \), demonstrating that the manifold is macroscopically degenerate while remaining microscopically nondegenerate. A complementary analysis of the first excitation gap for the frustrated lattices, presented in Fig.~\ref{fig.17}, further confirms the exponential suppression of the gap with increasing system size. In contrast, the unfrustrated chain retains a well-isolated ground state and a smoother low-energy spectrum.

In the \emph{strong-field regime} [Fig.~\ref{fig.gap}(b)], the transverse field dominates and suppresses the effects of geometric frustration. The ground state becomes unique and well separated from the excited states, and the overall spectrum of both geometries becomes smoother and more regularly spaced, eliminating the glass-like clustering observed under weak fields.

\subsection{RVB viewpoint and the challenge of qubit connectivity}
\label{subsec:RVB}

\begin{figure}[htbp]
    \centering
    \includegraphics[width=0.4\textwidth]{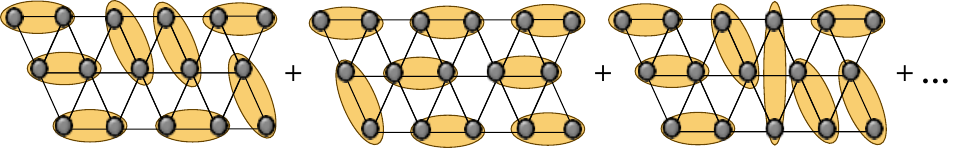}
    \caption{Superposition of possible valence-bond pairings in the RVB picture.}
    \label{fig:RVB} 
\end{figure}

\begin{figure}[htbp]
    \centering
    \includegraphics[width=0.2\textwidth]{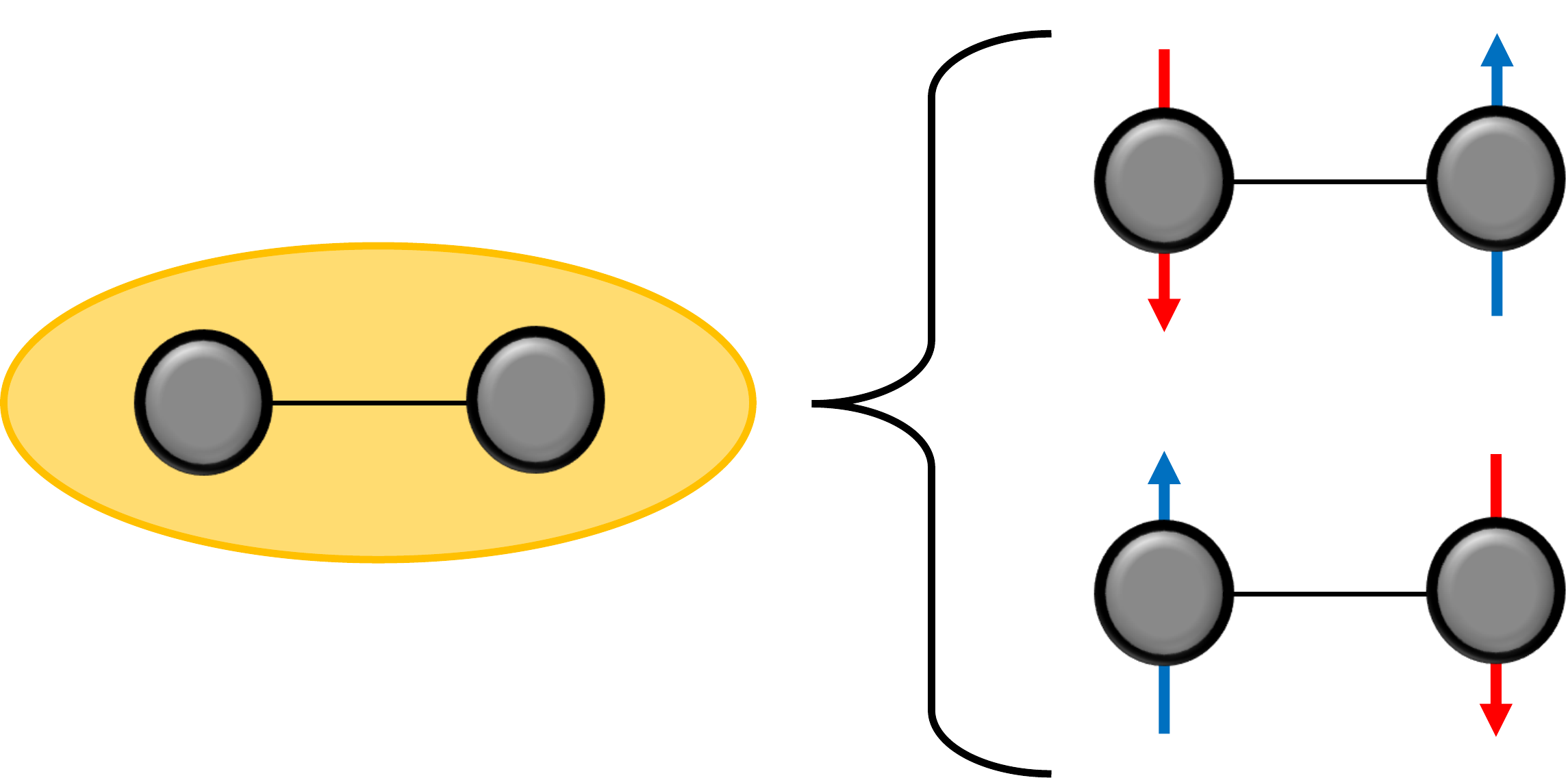}
   \caption{Schematic illustration of a singlet (VB) pair formed between two sites,
\( \ket{\psi_{\mathrm{pair}}}=\frac{1}{\sqrt{2}}
\bigl(\ket{\textcolor{red!90!black}{\downarrow}\textcolor{NavyBlue}{\uparrow}}
-\ket{\textcolor{NavyBlue}{\uparrow}\textcolor{red!90!black}{\downarrow}}\bigr) \).}
    \label{fig:pair} 
\end{figure}

In the broader context of frustrated magnetism, Anderson's
resonating-valence-bond (RVB) framework~\cite{anderson1973resonating} provides a
useful perspective. In Heisenberg-like spin models, where the Hamiltonian takes the form
\begin{equation}
    H_{\mathrm{Heis}} = \sum_{\langle i,j\rangle} J_{ij}\, 
    \mathbf{S}_i \cdot \mathbf{S}_j ,
\end{equation}
a singlet bond between two sites,
\(
\ket{\psi_{pair}} = \frac{1}{\sqrt{2}}
(\ket{\downarrow\uparrow}-\ket{\uparrow\downarrow}),
\)
forms an exact eigenstate of the two-site interaction and minimizes the local
energy. Accordingly, the ground state of the full system can be viewed as a
superposition of valence-bond coverings, illustrated in Fig.~\ref{fig:RVB},
with singlet (VB) pairs between sites forming the elementary building blocks of
the RVB framework (Fig.~\ref{fig:pair}).

Even for moderate system sizes, the number of distinct pairings grows combinatorially as $(2w)!/(2^w w!)$ for a system of $2w$ spins. While the RVB picture captures essential physics in Heisenberg-like models, directly importing this structure into a quantum circuit leads to practical difficulties, as encoding all possible pairing patterns results in a combinatorial explosion that would require exponentially many circuit components.

\begin{figure}[htbp]
    \centering
    \includegraphics[width=0.4\textwidth]{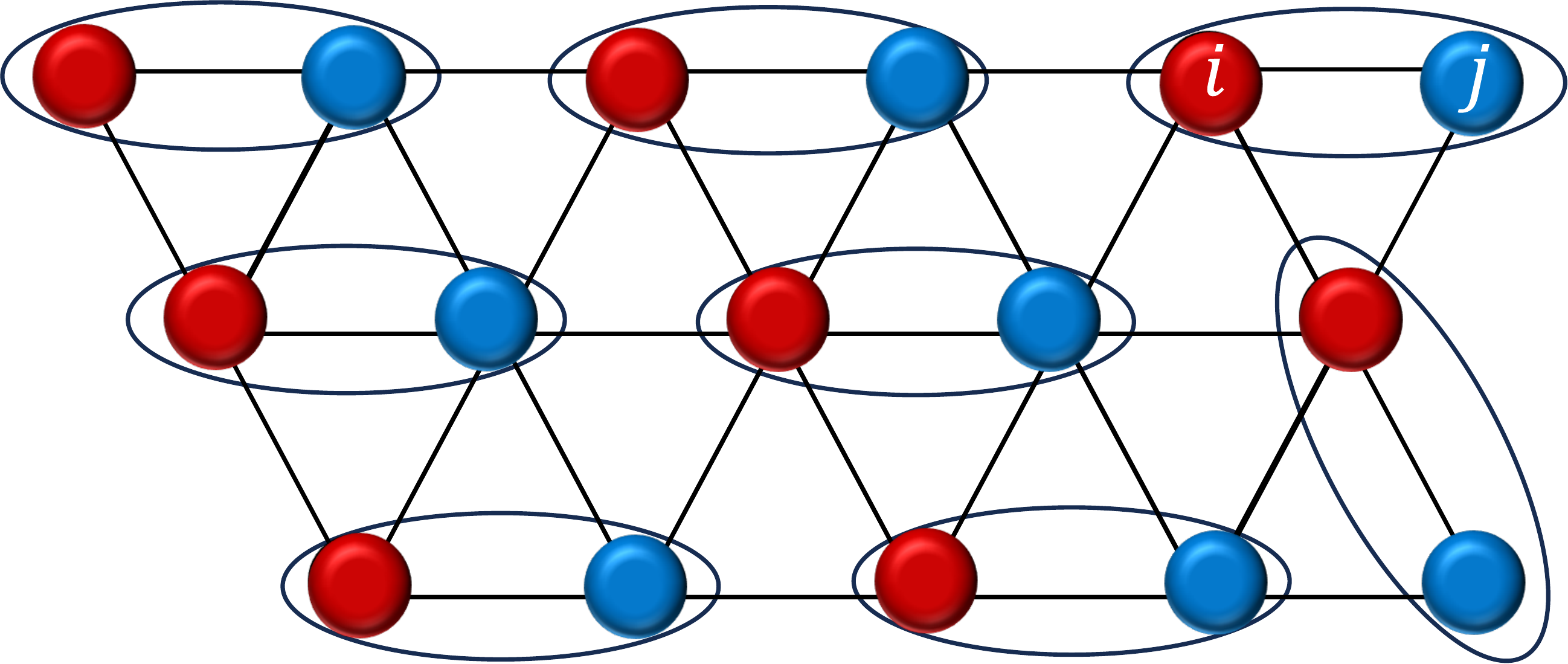}
    \caption{A triangular spin lattice colored by spin orientation, where red represents
spin-down \( \ket{\textcolor{red!90!black}{\downarrow}_i} \) and blue represents
spin-up \( \ket{\textcolor{NavyBlue}{\uparrow}_j} \). Edges connecting opposite 
    colors are satisfied; edges connecting identical colors are unsatisfied.}
    \label{fig.4} 
\end{figure}

A key observation is that the breakdown of a VB singlet into alternating spin-down/up components naturally resembles a \emph{bipartite coloring} of the underlying graph. Assigning red to spin-down and blue to spin-up partitions the lattice into two color classes, where an edge connecting opposite colors is \emph{satisfied} by the antiferromagnetic interaction. Conversely, edges connecting same-colored sites correspond to \emph{unsatisfied} bonds (Fig.~\ref{fig.4}). In this sense, each VB singlet implicitly enforces a local bipartite structure: no valence bond connects two sites of the same color. This coloring viewpoint provides a conceptual link between the RVB resonance picture and the cut-based classification of frustration introduced in the next section.

Thus, while the RVB analogy provides intuition about resonance in frustrated
systems, it is impractical as a direct design principle for constructing
shallow, hardware-efficient circuits for frustrated Ising models. In the next
section, we reinterpret the resonance concept from the RVB framework in a
different context, formulating an approach suited to the frustrated Ising
model.

\subsection{Frustration Classes and Cut-Based Resonance}
\label{subsec:frustration_classes}

In a geometrically frustrated Ising system, classical spin configurations can be
grouped according to how many interaction constraints they fail to satisfy. For
a configuration $C$, let $k(C)$ denote the number of unsatisfied AF edges, and
define the corresponding ``frustration class''
\begin{equation}
    \mathcal{C}_k = \{\, C \mid k(C) = k \,\}.
\end{equation}
This induces a natural hierarchy, where the lowest class $\mathcal{C}_0$
contains configurations with the minimal number of frustrated interactions, and
higher classes $\mathcal{C}_1, \mathcal{C}_2,\ldots$ contain progressively more
frustration. In the classical limit, the energy depends linearly on the number
of violated edges,
\begin{equation}
    E(C) \propto k(C),
\end{equation}
so the frustration classes form a well-defined energetic ordering.

When a transverse field is introduced, the ground state of the system is no
longer concentrated in a single classical configuration. The term
$a\sum_i X_i$ induces single-spin flips that mix different frustration classes,
leading to a quantum ground state $\ket{\psi_0}$ of the form
\begin{equation}
    \ket{\psi_0}
    = \sum_{k} \sum_{C\in\mathcal{C}_k} \alpha_C \ket{C},
\end{equation}
with amplitudes that decrease systematically with increasing frustration,
\begin{equation}
    |\alpha_{C\in\mathcal{C}_0}|
    > |\alpha_{C\in\mathcal{C}_1}|
    > |\alpha_{C\in\mathcal{C}_2}|
    > \cdots.
\end{equation}

Frustration classes admit a natural interpretation in terms of graph
partitioning: each classical spin configuration corresponds to a particular
partition (cut) of the interaction graph, and all configurations within a given
frustration class share the same number of edges that violate the AF constraint.
Thus, the hierarchy of frustration classes mirrors a hierarchy of graph cuts
ordered by the number of unsatisfied interactions.

Although the transverse-field Ising model does not exhibit singlet formation as
in SU(2)-symmetric Heisenberg models, geometric frustration induces an analogous
form of resonance. In Heisenberg systems, the ground state resonates among
multiple valence-bond coverings, whereas in the Ising case the resonance occurs
over the set of classical graph partitions compatible with the frustrated
geometry. However, the number of such partitions grows combinatorially, making
it inefficient to incorporate all of them directly into a quantum circuit.

To circumvent this challenge, we start with the ground state of the transverse field term in Eq.~(\ref{eq:TFIM}). This product state assigns equal weight to all computational-basis configurations, each representing a graph partition. We then use the dominant low-frustration partition to define which qubit pairs are connected by local two-qubit gates. These gates tune the configuration amplitudes and generate entanglement. This strategy captures the essential resonance structure of the frustrated system without requiring an explicit enumeration of all possible partitions.

This resonance structure naturally leads to a cut-based description. Minimizing
the number of unsatisfied edges corresponds directly to the Max-Cut problem.

\subsection{Max-Cut partitions}
\label{subsec:max-cut}
The goal of the Max-Cut problem is to divide the nodes of a graph into two distinct sets so as that maximize the edges that connect them. To label the two different groups, we can assign binary digits, 0 and 1. It is important to note that we cannot cut an edge that connects nodes within the same group. To solve the Max-Cut problem, we aim to maximize a specific cost function, which can be defined as follows
\begin{equation} 
 f(\vec x)= \\\sum_{i,j=0}^{n-1}w_{ij}x_i(1-x_j)\\,
 \label{eq.4}
\end{equation}
$w_{ij}$ represents the weight of the edge between nodes $i$ and $j$, and $\vec{x} = (x_{0}, x_{1}, \dots, x_{n-1}) \in \mathbb{B}^{n}$, where $\mathbb{B} = \{0, 1\}$ is a binary vector with a length corresponding to the number of nodes in the graph. Due to the graph being undirected, the weights are such that $w_{ij}$ is equal to $w_{ji}$.
Assigning a spin site $s_i$ to each node ($V$), where the spin value is $s_i=\pm1$, allows us to categorize the graph into two sets: Nodes with spin-up ($V^+$) and nodes with spin-down ($V^-$).  In the context of quantum computing, where we measure the expectation value of observables composed of Pauli operators, the connection to spin lattice mapping is logical and coherent. Indeed, this task can be accomplished by making the following assignment:
\begin{equation}
x_i=(1-Z_i)/2, 
\label{eq.5}
\end{equation}
where $Z_i$ denotes the Pauli $Z$ operator that has eigenvalues $\pm1$. This change of variable implies that
\begin{equation} 
\begin{split}
&x_i=0\rightarrow{} \: Z_i=1,\\
&x_i=1\rightarrow{} \: Z_i=-1 .
\end{split}
\label{eq.6}
\end{equation}
Upon employing Eq.~(\ref{eq.5}) and substituting it into Eq.~(\ref{eq.4}), we obtain
\begin{equation} 
 f(\vec z)= \\\sum_{i=0}^{n-1}\sum_{j=0}^{i}\frac{w_{ij}}{2}-\sum_{i=0}^{n-1}\sum_{j=0}^{i}\frac{w_{ij}}{2}Z_iZ_j.\\
\label{eq.7}
\end{equation}
Furthermore, the intrinsic inclination of a quantum computer is to seek minima, typically the lowest energy, rather than maxima. Therefore, instead of maximizing \(f(\vec{z})\), we aim to minimize
\begin{equation} 
 -f(\vec z)= \\\sum_{i=0}^{n-1}\sum_{j=0}^{i}\frac{w_{ij}}{2}Z_iZ_j-\sum_{i=0}^{n-1}\sum_{j=0}^{i}\frac{w_{ij}}{2}.\\
\label{eq.8}
\end{equation}
The observable for which we will calculate the expectation value is
\begin{equation} 
 H = \\\sum_{i=0}^{n-1}\sum_{j=0}^{i}\frac{w_{ij}}{2}Z_iZ_j,
\label{eq.9}
\end{equation}
with a constant correction.

In statistical physics and disordered systems, the Max-Cut problem is analogous to minimizing the Hamiltonian of an Ising spin glass without an external magnetic field. This reduces Eq.~(\ref{eq:TFIM}) to the following form
\begin{equation} 
 H = \\-\sum_{i,j}J_{ij}Z_iZ_j.\\
\label{eq.10}
\end{equation}
Utilizing the QAOA Algorithm \cite{farhi2014quantum} we can initially leverage a quantum computer to identify the maximum cut in an arbitrary graph. However, we use a more heuristic approach here, given the relatively small size of the system ($n \leq20$).

\subsection{Hamiltonian and symmetries}
\label{subsec:2}

It is straightforward to verify that the transverse-field Ising model Hamiltonian commutes with the total-spin flip operator
\begin{equation}
    X = \prod_{i=1}^NX_i.
    \label{eq.11}
\end{equation}
The proof boils down to showing that all local $Z_iZ_j$ commute with $X$.  We get 
\begin{align}
    [X,Z_iZ_j] &= \prod_{k\neq i,j}X_k[X_iX_j,Z_iZ_j]\nonumber\\
    & = \prod_{k\neq i,j}X_k \left(X_iZ_iX_jZ_j-Z_iX_iZ_jX_j\right)\nonumber\\
    & = \prod_{k\neq i,j}X_k \left((-iY_i)(-iY_j)-(iY_i)(iY_j)\right)\nonumber\\
    &=0 .
    \label{eq.12}
\end{align}
As a result, eigenstates of the Hamiltonian must also be eigenstates of the $X$ operator. Due to the idempotency of the $X$ operator ($X^2 = 1$), a $\mathbb{Z}_2$ symmetry arises, with eigenvalues $\pm1$ corresponding to the symmetric ($+$) and antisymmetric ($-$) sectors. The eigenstates of the $X$ operator exhibit a special property: any two states with the maximum Hamming distance (defined as the number of bit positions at which the corresponding binary strings differ) exhibit equal amplitude. However, their phases may vary. For example, in the two-qubit case, the eigenstates take the following form
\begin{equation}
    \ket{v^{\pm}}=\frac{1}{\sqrt{2}}\left(  \ket{00} \pm \ket{11} \right),
    \label{eq.13}
\end{equation}

\begin{equation}
    \ket{w^{^{\pm}}}=\frac{1}{\sqrt{2}}\left(  \ket{01} \pm \ket{10} \right).
    \label{eq.14}
\end{equation}
Consequently, a linear combination of symmetric states, $\ket{v^{+}} + \ket{w^{+}}$, or antisymmetric states, $\ket{v^{-}} + \ket{w^{-}}$, is also an eigenstate of the $X$ operator. Thus, the general form of the eigenstates can be rewritten as follows

\begin{equation}
    \ket{M^{\pm}}=\cos{x}\ket{v^{\pm}} + \sin{x} \ket{w^{\pm}}.
    \label{eq.15}
\end{equation}
Furthermore, $\ket{M^{\pm}}$ spans all states of the two-site Hamiltonian ($H_{2-\text{site}}$). Consider the two-site Hamiltonian
\begin{equation}
    H_{2-\text{site}}=bZ_1Z_0+a(X_{0} + X_{1}).
    \label{eq.16}
\end{equation}

The state $\ket{M^{+}}$ is the ground state in the second and fourth quadrants, with  
$\tan{\frac{x}{2}} = \frac{-b-\sqrt{ 4a^{2} - b^{2}} }{2a}$,  
and the highest excited state in the first and third quadrants, with  
$\tan{\frac{x}{2}} = \frac{-b+\sqrt{4a^{2} + b^{2}}}{2a}$. Thus we have
\begin{align}
    H_{2-\text{site}} \ket{M^{+}}& = \begin{cases}
+\sqrt{b^2 + 4a^2}\ket{M^{+}}, &  x \in  \text{QI or QIII} \\
-\sqrt{b^2 + 4a^2}\ket{M^{+}}, &  x \in \text{ QII or QIV}
\end{cases}
\label{eq.17}
\end{align}
Where QI–QIV denote the four quadrants. Similarly, $\ket{M^{-}}$ includes the first and second excited states, with $x = \frac{\pi}{2}$ yielding the first excited state $\ket{w^-}$ and $x = 0$ the second excited state $\ket{v^-}$.
 It is important to note that $\ket{M^{-}}$ is not an eigenstate of $H_{2\text{-}site}$ in general. However, it preserves the form of $\ket{M^{-}}$ through its action :
\begin{align}
    H_{2-\text{site}} \ket{M^{-}}& = b \big( \cos{x} \ket{v^{-}} - \sin{x}\ket{w^{-}}\big).
    \label{eq.18}
\end{align}
Preserving the form means the output state retains equal amplitudes for maximally Hamming distant states, with possible phase differences. This property arises from the commutation of the Ising Hamiltonian with the $X$ operator. By leveraging these properties, one can search for the target state using a quantum circuit within a confined searching space aligned with the target state's structure. This reduction is important as it addresses one of the primary sources of barren plateaus during cost landscape optimization \cite{ragone2024lie}. 

In the next section, we introduce the fundamental building block for designing a quantum circuit inspired by the symmetry of the system's Hamiltonian and and explain how these blocks are placed on the graph according to the Max-Cut partition.

\subsection{Max-Cut-partitioned ansatz}
\label{subsec:3}

Here, we present the fundamental circuit component for generating nearest-neighbor interactions, referred to as two-qubit cluster (2-qc).
\begin{equation}
\Qcircuit @C=0.7em @R=1em {
& \multigate{1}{\mathrm{2-qc}} & \qw\\
& \ghost{\mathbf{2-qc}}& \qw
}
\quad \raisebox{-0.8em}{$=$} \quad
\Qcircuit @C=0.7em @R=1em {
      & \ctrl{1} & \qw & \ctrl{1}  & \qw  \\
      & \targ & \gate{R_{y}(\theta)}  &  \targ  & \qw 
}\label{eq.19} 
\end{equation}  
This component is designed to align with the symmetry of the system's Hamiltonian and represents the partitioning mechanism using a cut for the simplest two-site system.
The 2-qc operator ($U_{2-qc}$) with qubit $i$ controlling $j$ can be written as 
\begin{equation}
    \texttt{CNOT}\cdot R_Y(2\theta)\cdot \texttt{CNOT}
    =\cos\theta I-i\sin\theta Y_jZ_i.
    \label{eq.20}
\end{equation}
Interestingly, the $U_{2-qc}$ commute with the $X$ operator $[X,U_{2-qc}] =0$, because,

\begin{align}
X_iX_j[\cos\theta I-i\sin\theta Y_jZ_i]X_iX_j =\cos\theta I-i\sin\theta Y_jZ_i.
\label{eq.21}
\end{align}
The consequence of this is that as soon as a reference state is an eigenstate of the $X$ operator
\begin{equation}
    X|\psi\rangle =(\pm)|\psi\rangle,
    \label{eq.22}
\end{equation}
acting with whatever $U_{2-qc}$ operator on the reference state will remain in the same symmetry sector.
\begin{equation}
    XU_{2-qc}\ket{\psi} = U_{2-qc}X\ket{\psi} =(\pm)U_{2-qc}\ket{\psi}.
    \label{eq.23}
\end{equation}

So, a circuit consisting of any 2-qc gates will be an eigenstate of the $X$ provided the reference state  (or initial state) $\prod_{i}\ket{q_{i}}$ is an eigenstate of $X$. The reference state $|\pm\rangle^{\otimes n}$ can be easily prepared as an eigenstate of the $X$ operator by applying an $R_y(\theta)$ gate to each qubit with $\theta = \pm \frac{\pi}{2}$.

\begin{lemma}
Let $U_{2\text{-}qc}$ denote the two-qubit cluster unitary that produces an exact eigenstate for the two-spin subsystem. The application of $U_{2\text{-}qc}$ does not alter the structure of $\ket{M^{\pm}}$; it merely shifts the effective rotation angle:
\begin{equation}
    U_{2\text{-}qc}\ket{M^{\pm}}
    =
    \cos\!\left(x \pm \tfrac{\theta}{2}\right)\ket{v^{\pm}}
    +
    \sin\!\left(x \pm \tfrac{\theta}{2}\right)\ket{w^{\pm}} .
    \label{eq:U2qc-action}
\end{equation}
\end{lemma}
\begin{proof}
The action of $U_{2\text{-}qc}$ is block-diagonal in the ordered basis
$\{\ket{00},\ket{10}\}\oplus\{\ket{11},\ket{01}\}$:
\[
U_{2\text{-}qc}
=
\begin{pmatrix}
\alpha & -\beta\\
\beta  & \alpha
\end{pmatrix}
\oplus
\begin{pmatrix}
\alpha & -\beta\\
\beta  & \alpha
\end{pmatrix},
\]
where
\[
\alpha=\cos\frac{\theta}{2},
\qquad
\beta=\sin\frac{\theta}{2}.
\]
Using the definitions of $\ket{v^\pm}$ and $\ket{w^\pm}$,
\begin{align}
U_{2\text{-}qc}\ket{v^\pm}
&=
\alpha\ket{v^\pm}
\pm\beta\ket{w^\pm},
\\
U_{2\text{-}qc}\ket{w^\pm}
&=
\mp\beta\ket{v^\pm}
+\alpha\ket{w^\pm}.
\end{align}
Therefore,
\begin{align}
U_{2\text{-}qc}\ket{M^\pm}
={}&
\left(
\alpha\cos x
\mp\beta\sin x
\right)\ket{v^\pm}
\nonumber\\
&+
\left(
\pm\beta\cos x
+\alpha\sin x
\right)\ket{w^\pm},
\end{align}
which reduces directly to Eq.~\eqref{eq:U2qc-action}.
\end{proof}

The next step is to generalize this circuit design to arbitrary graphs with $n$ nodes. 
This graph-based placement of 2-qc blocks, dictated entirely by a Max-Cut partition, 
defines our \emph{Max-Cut-Partitioned 2-QC Ansatz (MPA)}.

Consider a graph with nodes colored in two sets (``red'' and ``cyan'') according to a Max-Cut solution (Sec.~\ref{subsec:max-cut}). For each edge $(i,j)$ crossing the cut, there are two possible orientations for the 2-qc block, depending on whether $i$ or $j$ is chosen as the control (source, $s$) and which as the target (sink, $t$). For a single edge, this can be represented graphically as

\begin{equation}
    \centering
    \begin{minipage}{0.1\textwidth}
        \centering
        \begin{tikzpicture}
            \node[circle, fill=red!70, minimum size=0.7cm, font=\large , inner sep=2pt] (zero) at (0, 0) {0};
            \node[circle, fill=bluegreen!70, minimum size=0.7cm, font=\large,inner sep=2pt] (one) at (1.5, 0) {1};
            \draw[->, thick, dgreen] (0.5, 0.2) -- (1, 0.2);
            \draw[-,  line width=1.5pt] (zero) -- (one);
            \draw [decorate, decoration={brace, amplitude=10pt, raise=5pt}, yshift=-0.5cm]
            (-0.3, -0.4) -- (1.8, -0.4);
            \node at (0.75, -1.7) {%
                \Qcircuit @C=0.7em @R=1em {
                    \lstick{\ket{\textcolor{red}{q_0}}}  & \ctrl{1} & \qw & \ctrl{1}  &\qw  \\
                    \lstick{\ket{\textcolor{bluegreen}{q_1}}} &\targ & \gate{R_{y}(\theta)}  &  \targ  &\qw \\
                }
            };
        \end{tikzpicture}
    \end{minipage}
    \hspace{2cm}
    \begin{minipage}{0.1\textwidth}
        \centering
        \begin{tikzpicture}
            \node[circle, fill=red!70, minimum size=0.7cm, font=\large ,inner sep=2pt] (zero) at (0, 0.4) {0};
            \node[circle, fill=bluegreen!70, minimum size=0.7cm, font=\large ,inner sep=2pt] (one) at (1.5, 0.4) {1};
            \draw[<-, thick, dgreen] (0.5, 0.6) -- (1, 0.6);
            \draw[-,  line width=1.5pt] (zero) -- (one);
            \draw [decorate, decoration={brace, amplitude=10pt, raise=5pt}, yshift=-0.5cm]
            (-0.3, 0) -- (1.8, 0);
            \node at (0.75, -1.3) {%
                \raisebox{0.23cm}[\height][\depth]{%
                \Qcircuit @C=0.7em @R=1em {
            \lstick{\ket{\textcolor{red}{q_0}}}
            & \targ & \gate{R_y(\theta)} & \targ & \qw \\
            \lstick{\ket{\textcolor{bluegreen}{q_1}}}
            & \ctrl{-1} & \qw & \ctrl{-1} & \qw \\
        }
    }%
};
        \end{tikzpicture}
    \end{minipage}
\nonumber
\end{equation}

Given a Max-Cut coloring, we define two subcircuits:
\begin{itemize}
    \item the \emph{source} subcircuit $s$, where all 2-qc blocks are oriented such that all red nodes act as controls and cyan nodes as targets;
    \item the \emph{sink} subcircuit $t$, where the orientation is reversed: red nodes are targets, cyan nodes are controls.
\end{itemize}
The full ansatz layer is then built as an $s$–$t$ pair, preceded by single-qubit $R_y$ rotations for state preparation. The complete circuit for the triangle case is shown in Fig.~\ref{fig.6}. We also define a mean-field circuit or $\mathrm{cut}(0)$, which consists only of the initial $R_y$-layer without any 2-qc blocks; this reference is used to quantify the recovered correlation energy in Sec.~\ref{subsec:4}.

Two structural properties of the $s$ and $t$ subcircuits are worth highlighting:
\begin{enumerate}
    \item Within each subcircuit ($s$ or $t$), the 2-qc blocks commute with one another, since their rotations act either on distinct target qubits or on a shared target qubit with distinct control qubits. Consequently, the output state does not depend on the order in which the edges of a given cut are traversed, removing any combinatorial ambiguity in block placement.
    \item The relative order of $s$ and $t$ is unimportant: swapping the $s$ and $t$ layers amounts to a reparameterization and an overall shift of the cost landscape. In Sec.~\ref{sec:triangle_theorem} we explicitly show that both $s$–$t$ and $t$–$s$ orderings can produce the exact triangle ground state.
\end{enumerate}

By repeatedly applying the 2-qc blocks according to the Max-Cut coloring, we obtain a structured, symmetry-preserving ansatz for arbitrary graphs. In the next subsection, we specialize to a single triangle and show that, in this case, a single $s$–$t$ layer already has \emph{exact} expressibility for the ground state for all values of the parameters $a$ and $b$ in Eq.~(\ref{eq.16}).  

\subsection{Exact expressibility on the triangle}
\label{sec:triangle_theorem}

The frustrated triangle constitutes the minimal nontrivial system for
establishing our circuit design methodology. The analytical results for this
case are summarized below as a theorem, highlighting the central claim of this
section.

\begin{theorem}
Consider the three-site antiferromagnetic transverse-field Ising Hamiltonian on a triangle with uniform couplings $J_{ij}=-b$ and transverse field $h=-a$ as in Eq.~(\ref{eq:TFIM}). Let $\ket{\psi_{\mathrm{tgt}}}$ denote its ground state. Then for all $a,b\ge 0$ there exists a single real parameter $x^\star$ such that one layer of the Max-Cut-partitioned 2-qc circuit (one $s$–$t$ pair, Fig.~\ref{fig.6}) prepares the exact ground state,
\begin{equation}
    \ket{\psi_{s\text{-}t}(x^\star)} = \ket{\psi_{\mathrm{tgt}}}.
\end{equation}
Equivalently, the ansatz reproduces both the exact ground-state energy
\begin{equation}
    E_{\text{tgt}} = -a + b - 2\sqrt{a^2 + ab + b^2},
\end{equation}
and the symmetry-adapted amplitude structure of $\ket{\psi_{\mathrm{tgt}}}$ using a single variational angle.
\end{theorem}

\begin{figure*}[!htbp]
\centering
\hspace{2.7em}
\begin{tikzpicture}

    \tikzstyle{node0} = [circle, draw, fill=red, minimum size=0.7cm, inner sep=2pt]
    \tikzstyle{node1} = [circle, draw, fill=bluegreen!70, minimum size=0.7cm, inner sep=2pt]

    \node[node0] (0) at (0, 0) {0};
    \node[node1] (1) at (2, 0) {1};
    \node[node1] (2) at (1, 1.732) {2}; 

    \draw[line width=1.5pt] (0) -- (1);
    \draw[line width=1.5pt](1) -- (2);
    \draw[line width=1.5pt] (2) -- (0);

    \draw[->, thick, dgreen] (0.03, 0.6) -- (0.5, 1.3);
    \draw[->, thick, dgreen] (0.6, -0.2) -- (1.5, -0.2);
        \draw[dashed,line width=1pt] (-0.5, 0.4) .. controls (0.5, 0.5) .. (0.5, -0.3);
\end{tikzpicture}
\hspace{8em}
\begin{tikzpicture}

    \tikzstyle{node0} = [circle, draw, fill=red, minimum size=0.7cm, inner sep=2pt]
    \tikzstyle{node1} = [circle, draw, fill=bluegreen!70, minimum size=0.7cm, inner sep=2pt]

    \node[node0] (0) at (0, 0) {0};
    \node[node1] (1) at (2, 0) {1};
    \node[node1] (2) at (1, 1.732) {2}; 

    \draw[line width=1.5pt] (0) -- (1);
    \draw[line width=1.5pt](1) -- (2);
    \draw[line width=1.5pt] (2) -- (0);

    \draw[<-, thick, dgreen] (0.03, 0.6) -- (0.5, 1.3);
    \draw[<-, thick, dgreen] (0.6, -0.2) -- (1.5, -0.2);
        \draw[dashed,line width=1pt] (-0.5, 0.4) .. controls (0.5, 0.5) .. (0.5, -0.3);
\end{tikzpicture}
\Qcircuit @C=0.7em @R=1em { \\
& & & & & & \mbox{$\boldsymbol{s}$} & & & & & & & & & \mbox{$\boldsymbol{t}$}& & & & & & \\
    \lstick{\ket{\textcolor{red}{q_0}}} & \gate{R_{y}(\theta_{0})} & \qw & \ctrl{1} & \qw & \ctrl{1} & \qw & \ctrl{2} & \qw & \ctrl{2} & \qw & \qw & \targ & \gate{R_{y}(\theta_{5})} & \targ & \qw & \targ & \gate{R_{y}(\theta_{6})} & \targ & \qw & \qw \\
    \lstick{\ket{\textcolor{bluegreen}{q_1}}} & \gate{R_{y}(\theta_{1})} & \qw & \targ & \gate{R_{y}(\theta_{3})} & \targ & \qw & \qw & \qw & \qw & \qw & \qw & \ctrl{-1} & \qw & \ctrl{-1} & \qw & \qw & \qw & \qw & \qw & \qw \\
    \lstick{\ket{\textcolor{bluegreen}{q_2}}} & \gate{R_{y}(\theta_{2})} & \qw & \qw & \qw & \qw & \qw & \targ & \gate{R_{y}(\theta_{4})} & \targ & \qw & \qw & \qw & \qw & \qw & \qw & \ctrl{-2} & \qw & \ctrl{-2} & \qw & \qw \gategroup{3}{4}{5}{10}{1.4em}{--} 
    \gategroup{3}{13}{5}{19}{1.4em}{--} 
}
    \caption{The whole circuit for the AF Ising spins on a triangle}\label{fig.6} 
\end{figure*}

\begin{proof}[Proof]
To prove the claim, we analyze the symmetry structure of the triangular Ising Hamiltonian. The triangular Ising model has $2^3 = 8$ possible configurations. We can categorize the possible configurations based on the number of unsatisfied edges, where both sites on an edge have the same spin. In total, there are two different classes of configurations: those with one unsatisfied edge (1U class) and those with three unsatisfied edges (3U class). The latter case represents a less favorable state due to its higher energy. The general wavefunction has the following form:
\begin{equation}
    \ket{\psi}=N(\sum_{i=0}^{1}p_{i}\ket{P_{i}} + \sum_{j=0}^{5}q_{j}\ket{Q_{j}}),
\end{equation}

\begin{table}[htbp]
    \centering
    \caption{Labeling scheme for $\ket{P_{i}}$ and $\ket{Q_{i}}$ states, categorized into 3U and 1U classes.
}
    \label{tab.1}
    \begin{tabular}{c c c}
        \hline
        Class & Label & Basis State \\ 
        \hline
        \multirow{2}{*}{3U} & $\ket{P_{0}}$ & $\ket{000}$ \\ 
                                       & $\ket{P_{1}}$ & $\ket{111}$ \\ 
        \hline
        \multirow{6}{*}{1U} & $\ket{Q_{0}}$ & $\ket{001}$ \\ 
                                       & $\ket{Q_{1}}$ & $\ket{010}$ \\ 
                                       & $\ket{Q_{2}}$ & $\ket{011}$ \\ 
                                       & $\ket{Q_{3}}$ & $\ket{100}$ \\ 
                                       & $\ket{Q_{4}}$ & $\ket{101}$ \\ 
                                       & $\ket{Q_{5}}$ & $\ket{110}$ \\ 
        \hline
    \end{tabular}
\end{table}

where $N$ is the normalization factor, $\ket{P_{i}}$  and $\ket{Q_{j}}$ represent 3U and 1U classes respectively. $p_i$ and $q_j$ represent the amplitudes corresponding to configurations in each class. Those configurations in the same class have equal amplitudes in terms of magnitude. This equality results from their symmetry and balanced contribution to the wavefunction, where none of the configurations is more stable than the others, making them equally significant. However, the phase of the amplitudes for the same class can vary. We can denote the magnitudes of amplitudes for classes 3U and 1U as $\cos{\theta}$ and $\sin{\theta}$, respectively, leveraging their equivalence in absolute amplitudes within the same class. Consequently, the normalization factor can be determined as follows:
\begin{equation}
	|p_0|=|p_1|=\cos{\theta} \ \ \ , \ \ \  |q_0|=|q_1|=...|q_5|=\sin{\theta},\\
\end{equation}
\begin{gather}
	N=\frac{1}{\sqrt{2\left(2-\cos{2\theta}\right)}}.
\end{gather}

Similar to the two-site case, we can demonstrate the ground state based on the eigensates of the $X$ operator using the labeling scheme in Table \ref{tab.1}. The eight eigenstates of the $X$ operator are:
\begin{equation}
\ket{v_{0}^{\pm}}= \frac{1}{\sqrt{2}}\left(\ket{P_0}\pm \ket{P_1}  \right),
\end{equation}

\begin{equation}
\ket{v_{1}^{\pm}}= \frac{1}{\sqrt{2}}\left(\ket{Q_5}\pm \ket{Q_0}\right),
\end{equation}

\begin{equation}
\ket{v_{2}^{\pm}}= \frac{1}{\sqrt{2}}\left(\ket{Q_4}\pm \ket{Q_1}\right),
\end{equation}

\begin{equation}
\ket{v_{3}^{\pm}}= \frac{1}{\sqrt{2}}\left(\ket{Q_2}\pm \ket{Q_3}\right).
\end{equation}

The basis set $\{\ket{v_{i}^{\pm}}\}$ spans all states of the Hamiltonian. The proper phase for each basis state in constructing the ground state is determined by examining its expectation energy in the general form across all possible phase assignments. We omit these details for brevity and instead focus on the ground state, which serves as our target state.
 It can be shown explicitly that the target state $\ket{\psi_{\textrm{tgt}}}$ has the following form
\begin{equation}
    \ket{\psi_{\textrm{tgt}}}=\frac{1}{\sqrt{ 2-\cos{2\theta} }}\left( \cos{\theta}\ket{v_{0}^{-}} - \sin{\theta} \sum_{i=1}^{3}\ket{v_{i}^{-}} \right),
    \label{eq.33}
\end{equation}
where the angle $\theta$ satisfies
\begin{equation}
    \tan \theta = \frac{(2b + a) + 2\sqrt{a^2 + ab + b^2}}{3a},
\end{equation}
and the ground-state energy is given by
\begin{equation}
    E_{\text{tgt}} = -a + b - 2\sqrt{a^2 + ab + b^2}.
\end{equation}
 
Each configuration in the target state is characterized by a sign determined by the number of spin-up and spin-down sites. Configurations with an odd number of 1's are assigned a minus sign, whereas those with an even number of 1's have a plus sign. For instance, $\ket{001}$ holds a minus sign, whereas $\ket{110}$ carries a plus sign. Therefore, if two different configurations have a maximum Hamming distance, they should exhibit opposite signs with respect to each other.

As we can see, the target state consists of two coefficients, implying that only one parameter is required to obtain it. To maintain consistency, the circuit in Fig~\ref{fig.6} reproduces the desired state using a single parameter, for all $a$ and $b$ values as shown below.
\begin{widetext}
\begin{align}
\Qcircuit @C=0.7em @R=1em {
    \lstick{\ket{\textcolor{red}{q_0}}} & \gate{R_{y}(-\frac{\pi}{2})} & \qw \barrier{2}  & \ctrl{1} & \qw & \ctrl{1}  &\qw & \ctrl{2} & \qw &\ctrl{2} & \qw \qw \barrier{2} & \targ & \gate{R_{y}(-\frac{\pi}{2})}  &\targ & \targ & \gate{R_{y}(\frac{x}{2})}  &\targ &\qw\\
    \lstick{\ket{\textcolor{bluegreen}{q_1}}} & \gate{R_{y}(-\frac{\pi}{2})} & \qw   &\targ & \gate{R_{y}(\frac{y}{2})}  &  \targ  &\qw & \qw & \qw &\qw &\qw  & \ctrl{-1} &\qw &\ctrl{-1} &\qw &\qw &\qw&\qw\\
    \lstick{\ket{\textcolor{bluegreen}{q_2}}} & \gate{R_{y}(-\frac{\pi}{2})}  & \qw  &  \qw & \qw & \qw & \qw & \targ  &\gate{R_{y}(\frac{x}{2})} &\targ  & \qw  & \qw & \qw & \qw& \ctrl{-2} &\qw &\ctrl{-2}&\qw\\
}
\end{align}
\end{widetext}
where $\tan(\frac{y}{2}) = \sqrt{2}\sin{(x + \pi/4)})$ , resulting in the output state:
\begin{equation}
    \psi_{s-t}= C \bigg(
        \left(2\sin{x}+\cos{x} \right) \ket{v_{0}^{-}} 
         +\cos{x} \sum_{i=1}^{3}\ket{v_{i}^{-}} 
    \bigg),
\end{equation}
here ${s-t}$ represents a single layer of  subcircuit $s$ and subcircuit $t$, with the order of $s$ and $t$ proceeding from left to right and $C=\frac{1}{\sqrt{2\left(2+\sin(2x)\right)}}$. By applying the identity $\tan x = \frac{1 - \cot{\theta}}{2}$, one can derive Eq.~(\ref{eq.33}). In this context, the variable $x$ satisfies the following relation:
\begin{equation}
    \tan x = \frac{-(a + b) + \sqrt{a^2 + ab + b^2}}{a}.
\end{equation}

As mentioned earlier, we can swap subcircuits $s$ and $t$ to obtain the target state, and we derive:
\begin{widetext}
\begin{align}
\Qcircuit @C=0.7em @R=1em { \lstick{\ket{\textcolor{red}{q_0}}} & \gate{R_{y}(-\frac{\pi}{2})} & \qw \barrier{2} & \targ & \gate{R_{y}(\frac{y'}{2})} & \targ &\qw & \targ &  \gate{R_{y}(\frac{x'}{2})} & \targ & \qw \qw \barrier{2} & \ctrl{1} & \qw & \ctrl{1}  &\ctrl{2} & \qw&\ctrl{2}  & \qw\\ \lstick{\ket{\textcolor{bluegreen}{q_1}}} & \gate{R_{y}(-\frac{\pi}{2})} & \qw & \ctrl{-1} & \qw & \ctrl{-1} & \qw & \qw & \qw & \qw & \qw & \targ & \gate{R_{y}(\frac{\pi}{2})} & \targ & \qw & \qw &\qw & \qw\\ \lstick{\ket{\textcolor{bluegreen}{q_2}}} & \gate{R_{y}(-\frac{\pi}{2})} & \qw & \qw & \qw & \qw & \qw & \ctrl{-2} & \qw & \ctrl{-2} & \qw & \qw & \qw & \qw& \targ & \gate{R_{y}(\frac{x'}{2})}  & \targ & \qw \\ 
}
\end{align}
\end{widetext}
where $y' = 2\tan^{-1}(\sqrt{2}\sin{(x' - \frac{\pi}{4})})$, resulting in the output state:
\begin{equation}
    \psi_{t-s} = C' \bigg(
        \sin{(x'+\frac{\pi}{4}))} \ket{v_{0}^{-}}
         +\cos{(x'+\frac{\pi}{4}))}\sum_{i=1}^{3}\ket{v_{i}^{-}} 
    \bigg),
\end{equation}
here $C'= \frac{1}{\sqrt{2-\sin(2x')}}$. By using the identity $\tan\left(x' + \frac{\pi}{4}\right) = -\cot\theta$, Eq.~(\ref{eq.33}) can be derived. This leads to the following relationship between $x'$ and the parameters $a$ and $b$ of the Hamiltonian:
\begin{equation}
    \tan x' = \frac{-(a + 2b) + 2\sqrt{a^2 + ab + b^2}}{a}.
\end{equation}
\end{proof}
For larger systems, we will apply the 2-qc building block to benchmark the trainability of the circuit in finding the target state. The capability of this framework to find the target state for larger frustrated systems is explored in the following section. Whereas the MPA was able to reproduce the ground state for the triangle exactly, this is no longer the case for larger graphs. However, taking the triangle as building blocks for frustrated systems, one can employ the $s$ and $t$ circuits in the same vein.

\section{Numerical methodology}
\label{sec:methods}

\subsection{Extracting Max-Cut bitstrings}
\label{subsec:maxcut_finding}

To obtain the Max-Cut, we use Eq.~(\ref{eq.10}) and set $J_{ij}=1$
to assign equal weight to all edges in the graph.
We run the QAOA algorithm and identify the bitstrings with the highest
output probabilities, which correspond to the Max-Cut solutions.

\subsection{Performance metrics}
\label{subsec:metrics}

We assess the performance of the MPA method using complementary metrics that probe both
energetic accuracy and the quality of the prepared quantum state.

\subsubsection{Recovered correlation energy}

The recovered correlation energy (RCE) quantifies how much of the correlation
energy—defined as the energy difference between the exact ground state and the
mean-field approximation~\cite{szabo1989modern}—is captured by the MPA. It is computed as
\begin{equation}
    \% \mathrm{RCE}
    =
    \frac{E_{\mathrm{mf}} - E_{\mathrm{MPA}}}{E_{\mathrm{mf}} - E_{\mathrm{tgt}}}
    \times 100 ,
\end{equation}
where \(E_{\mathrm{mf}}\) is the mean-field energy obtained from a circuit containing
only single-qubit \(R_y\) rotations (i.e., without any two-qubit entangling blocks),
\(E_{\mathrm{MPA}}\) denotes the energy obtained using the MPA circuit, and \(E_{\mathrm{tgt}}\) is the target-state energy. Negative values of $\%\mathrm{RCE}$ indicate that the variational ansatz performs worse than the mean-field reference, i.e., it fails to recover correlation energy beyond a product-state description.

\subsubsection{Fidelity}

To quantify the closeness of the MPA state to the target state, we use the Uhlmann
fidelity~\cite{Nielsen,uhlmann1976transition}. For the MPA density matrix
\(\rho_{\mathrm{MPA}}\) and the target density matrix \(\rho_{\mathrm{tgt}}\), the
fidelity is defined as
\begin{equation}
    F(\rho_{\mathrm{MPA}}, \rho_{\mathrm{tgt}})
    =
    \left[
        \mathrm{Tr}\!\left(
            \sqrt{
                \sqrt{\rho_{\mathrm{MPA}}}\,
                \rho_{\mathrm{tgt}}\,
                \sqrt{\rho_{\mathrm{MPA}}}
            }
        \right)
    \right]^2 .
\end{equation}

We note that fidelity is not a directly observable quantity on current quantum hardware, as its estimation requires either access to the exact target state or full (or partial) quantum state tomography. Both approaches are highly sensitive to statistical shot noise and scale unfavorably with system size. For this reason, fidelity is evaluated here using ideal simulations with infinite shots, where it serves as a diagnostic tool to assess the intrinsic expressibility of the variational ansatz rather than as a hardware-level benchmark.

\subsubsection{Correlation functions and magnetization}

Although energy and fidelity provide good global measures of performance, we also
employ experimentally accessible observables—namely correlation functions
$\langle Z_i Z_j\rangle$ and magnetizations $\langle X_i\rangle$—as performance
metrics for the 19- and 20-site lattices.

\subsection{Simulation setup}
\label{subsec:sim_setup}

All simulations were performed using the Qiskit Aer \texttt{Estimator} in the state vector mode, under noiseless conditions~\cite{Qiskit}. The classical optimisation was carried out using SLSQP~\cite{kraft1988software} with a convergence threshold of \(10^{-8}\). The performance of the proposed method was benchmarked against QAOA~\cite{farhi2014quantum} and ADAPT-VQE~\cite{grimsley2019adaptive}, with a detailed description of these approaches provided in Appendix~\ref{a1}. For a fair comparison, QAOA and ADAPT-VQE simulations were run using identical classical optimization settings. In all cases, variational parameters were initialized randomly using a single initialization per simulation. Additional hardware-realistic simulations, incorporating a
backend-derived noise model, are reported separately in
Appendix~\ref{app:noise} and are restricted to energy-based observables.

\subsection{Optimization protocol}
\label{subsec:optim_protocol}

Parameters are optimised across a grid of \(a\in[0,1]\).  For large
graphs (19–20 sites), an adiabatic continuation scheme is used: the
optimal parameters at one value of \(a\) are used as the initial point
for the next value.

\section{Result and discussion}
\label{sec:methods}
\subsection{5-site frustrated Ising model}
\label{subsec:4}

Here, we present results for a 5-site frustrated Ising model on the house graph \cite{wiki:maximumcut}. For various cuts, including Max Cut to evaluate the circuit's performance across different graph partitioning approaches (see Fig.~\ref{fig.7}). To distinguish between different cuts, we define $\mathrm{cut}(\alpha)$, where $\alpha$ denotes the number of edges in a cut. In this case, the Max-Cut corresponds to $\mathrm{cut}(5)$. In each scenario, the circuit starts from the initial state and employs three layers of $s$, $t$, and $s$ subcircuits in the order of $s - t - s$. The resulting recovered correlation energy and state fidelity are presented in Fig.~\ref{fig.8}. 

\begin{figure}[htbp]
  \centering
  \subfloat[cut(2)]{
    \begin{tikzpicture}[scale=0.25, 
    every node/.style={circle, draw, minimum size=0.2cm, font=\bfseries},
    plain/.style={draw=none, fill=none, font=\bfseries},
    baseline=(current bounding box.north)
]

\node[fill=bluegreen!70] (n0) at (0, 0) {0};
\node[fill=bluegreen!70] (n1) at (6,0) {1};
\node[fill=red!70]  (n2) at (0,6) {2};
\node[fill=red!70]  (n3) at (6,6) {3};
\node[fill=red!70]  (n4) at (3,10) {4};

\draw (n0) -- (n1);
\draw (n0) -- (n2);
\draw (n2) -- (n3);
\draw (n1) -- (n3);
\draw (n2) -- (n4);
\draw (n3) -- (n4);

\draw[dashed, ultra thick, black]
  (-2,1) .. controls (3,3)..(8,1);

\end{tikzpicture}
}
\subfloat[cut(2*)]{
    \begin{tikzpicture}[scale=0.25, 
    every node/.style={circle, draw, minimum size=0.2cm, font=\bfseries},
    plain/.style={draw=none, fill=none, font=\bfseries},
    baseline=(current bounding box.north)
]

\node[fill=red!70
] (n0) at (0, 0) {0};
\node[fill=bluegreen!70] (n1) at (6,0) {1};
\node[fill=red!70]  (n2) at (0,6) {2};
\node[fill=red!70]  (n3) at (6,6) {3};
\node[fill=red!70]  (n4) at (3,10) {4};

\draw (n0) -- (n1);
\draw (n0) -- (n2);
\draw (n2) -- (n3);
\draw (n1) -- (n3);
\draw (n2) -- (n4);
\draw (n3) -- (n4);

\draw[dashed, ultra thick, black]
  (5,-2) .. controls (4,2)..(8,1);


\end{tikzpicture}
}
\\
\subfloat[cut(3)]{
    \begin{tikzpicture}[scale=0.25, 
    every node/.style={circle, draw, minimum size=0.2cm, font=\bfseries},
    plain/.style={draw=none, fill=none, font=\bfseries},
    baseline=(current bounding box.north)
]

\node[fill=bluegreen!70] (n0) at (0, 0) {0};
\node[fill=red!70] (n1) at (6,0) {1};
\node[fill=bluegreen!70]  (n2) at (0,6) {2};
\node[fill=red!70]  (n3) at (6,6) {3};
\node[fill=red!70]  (n4) at (3,10) {4};

\draw (n0) -- (n1);
\draw (n0) -- (n2);
\draw (n2) -- (n3);
\draw (n1) -- (n3);
\draw (n2) -- (n4);
\draw (n3) -- (n4);

\draw[dashed, ultra thick, black]
  (1,-2) .. controls (3,5)..(0.5,10);


\end{tikzpicture}
}
\subfloat[cut(3*)]{
    \begin{tikzpicture}[scale=0.25, 
    every node/.style={circle, draw, minimum size=0.2cm, font=\bfseries},
    plain/.style={draw=none, fill=none, font=\bfseries},
    baseline=(current bounding box.north)
]

\node[fill=red!70] (n0) at (0, 0) {0};
\node[fill=red!70] (n1) at (6,0) {1};
\node[fill=bluegreen!70]  (n2) at (0,6) {2};
\node[fill=red!70]  (n3) at (6,6) {3};
\node[fill=red!70]  (n4) at (3,10) {4};

\draw (n0) -- (n1);
\draw (n0) -- (n2);
\draw (n2) -- (n3);
\draw (n1) -- (n3);
\draw (n2) -- (n4);
\draw (n3) -- (n4);

\draw[dashed, ultra thick, black]
  (-2,5) .. controls (2,4)..(1,9);


\end{tikzpicture}
}
\\
 \subfloat[cut(4)]{
    \begin{tikzpicture}[scale=0.25, 
    every node/.style={circle, draw, minimum size=0.2cm, font=\bfseries},
    plain/.style={draw=none, fill=none, font=\bfseries},
    baseline=(current bounding box.north)
]

\node[fill=red!70] (n0) at (0, 0) {0};
\node[fill=red!70] (n1) at (6,0) {1};
\node[fill=bluegreen!70]  (n2) at (0,6) {2};
\node[fill=bluegreen!70]  (n3) at (6,6) {3};
\node[fill=red!70]  (n4) at (3,10) {4};

\draw (n0) -- (n1);
\draw (n0) -- (n2);
\draw (n2) -- (n3);
\draw (n1) -- (n3);
\draw (n2) -- (n4);
\draw (n3) -- (n4);

\draw[dashed, ultra thick, black]
  (-1,10) .. controls (2,7)..(7,8)
  .. controls (9,6)..(7,4)
  .. controls (3,4)..(-2,2)
  
  ;


\end{tikzpicture}
}
 \subfloat[cut(4*)]{
    \begin{tikzpicture}[scale=0.25, 
    every node/.style={circle, draw, minimum size=0.2cm, font=\bfseries},
    plain/.style={draw=none, fill=none, font=\bfseries},
    baseline=(current bounding box.north)
]

\node[fill=red!70] (n0) at (0, 0) {0};
\node[fill=bluegreen!70] (n1) at (6,0) {1};
\node[fill=red!70]  (n2) at (0,6) {2};
\node[fill=red!70]  (n3) at (6,6) {3};
\node[fill=bluegreen!70]  (n4) at (3,10) {4};

\draw (n0) -- (n1);
\draw (n0) -- (n2);
\draw (n2) -- (n3);
\draw (n1) -- (n3);
\draw (n2) -- (n4);
\draw (n3) -- (n4);

\draw[dashed, ultra thick, black]
  (-1,10) .. controls (2,7)..(7,8)
  .. controls (9,6)..(7,4)
  .. controls (5,4)..(3,-2)
  
  ;

\end{tikzpicture}
}
\\
\subfloat[cut(5) = Max-Cut]{
    \begin{tikzpicture}[scale=0.25, 
    every node/.style={circle, draw, minimum size=0.2cm, font=\bfseries},
    plain/.style={draw=none, fill=none, font=\bfseries},
    baseline=(current bounding box.north)
]

\node[fill=bluegreen!70] (n0) at (0, 0) {0};
\node[fill=red!70] (n1) at (6,0) {1};
\node[fill=red!70]  (n2) at (0,6) {2};
\node[fill=bluegreen!70]  (n3) at (6,6) {3};
\node[fill=bluegreen!70]  (n4) at (3,10) {4};

\draw (n0) -- (n1);
\draw (n0) -- (n2);
\draw (n2) -- (n3);
\draw (n1) -- (n3);
\draw (n2) -- (n4);
\draw (n3) -- (n4);

\draw[dashed, ultra thick, black]
  (0,10) .. controls (2,5)..(7,1)
  .. controls (9,-3)..(-3,3)
   ;

\end{tikzpicture}
}
  \caption{Six possible cuts for the 5-site house configuration, including the Max-Cut.
 }
  \label{fig.7}
\end{figure}

Fig.~\ref{fig.8} demonstrates that the Max-Cut–based MPA circuit achieves the exact solution with high fidelity across the entire interval of \(a\). However, for cut(4) and cut(4*), the exact result is reproduced only when \(a \geq 0.5\), and as \(a\) decreases, the fidelity drops rapidly.
 As previously stated, simulating spin glasses becomes particularly challenging when the magnetic field is weak. Consequently, if the circuit provides an incomplete representation compared to the target state, it will readily align with the structure of an excited state. This leads to convergence to a linear combination of the ground state and the excited state, rather than the ground state itself.
 
\begin{figure}[htbp]
    \centering
    \subfloat[]{%
        \includegraphics[width=0.4\textwidth]{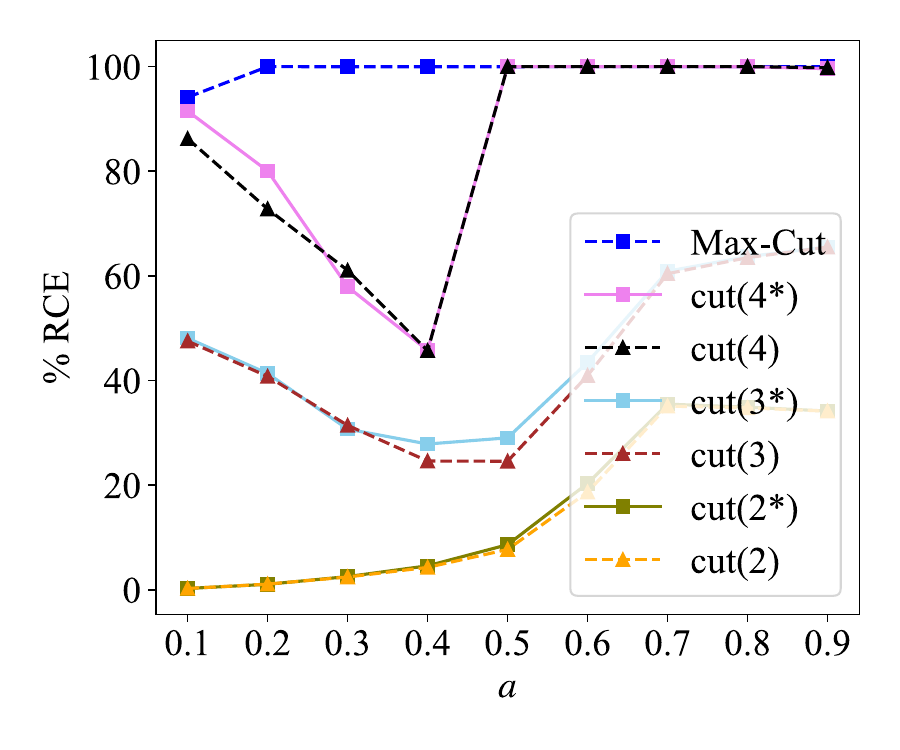}%
        \label{fig.8(a)}%
    } \hspace{0.06\textwidth}
    \subfloat[]{%
        \includegraphics[width=0.4\textwidth]{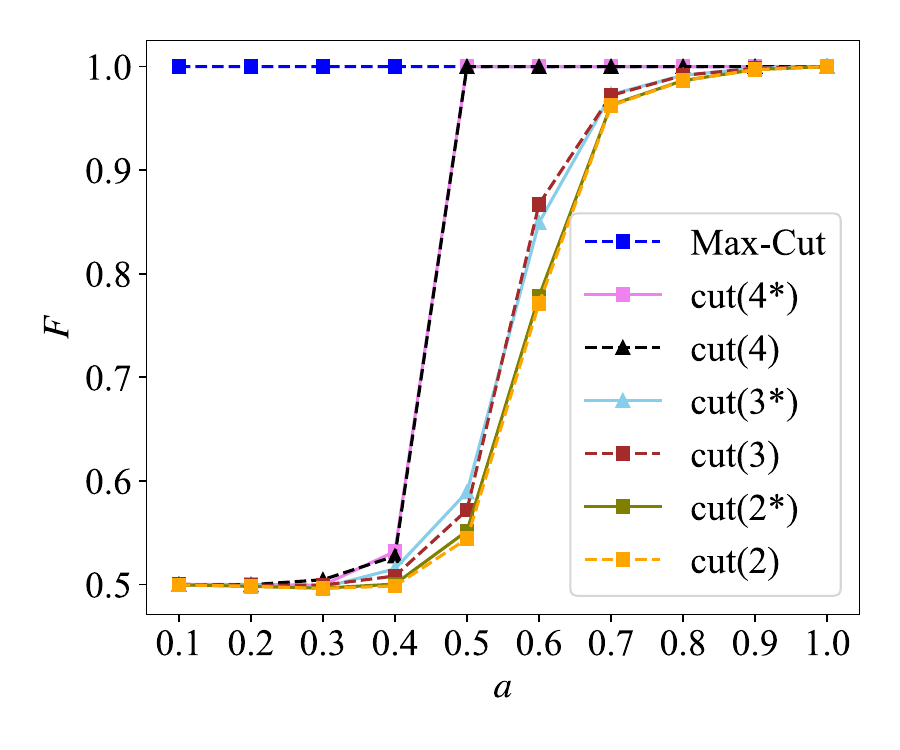}%
        \label{fig.8(b)}%
    }
    \caption{Comparison of the recovered correlation energy for different cuts, including the Max-Cut (a), and the state fidelity (b) for the 5-site house frustrated Ising model.}
    \label{fig.8}
\end{figure}

\begin{figure}[htbp]    
    \centering
    \includegraphics[width=0.5\textwidth]{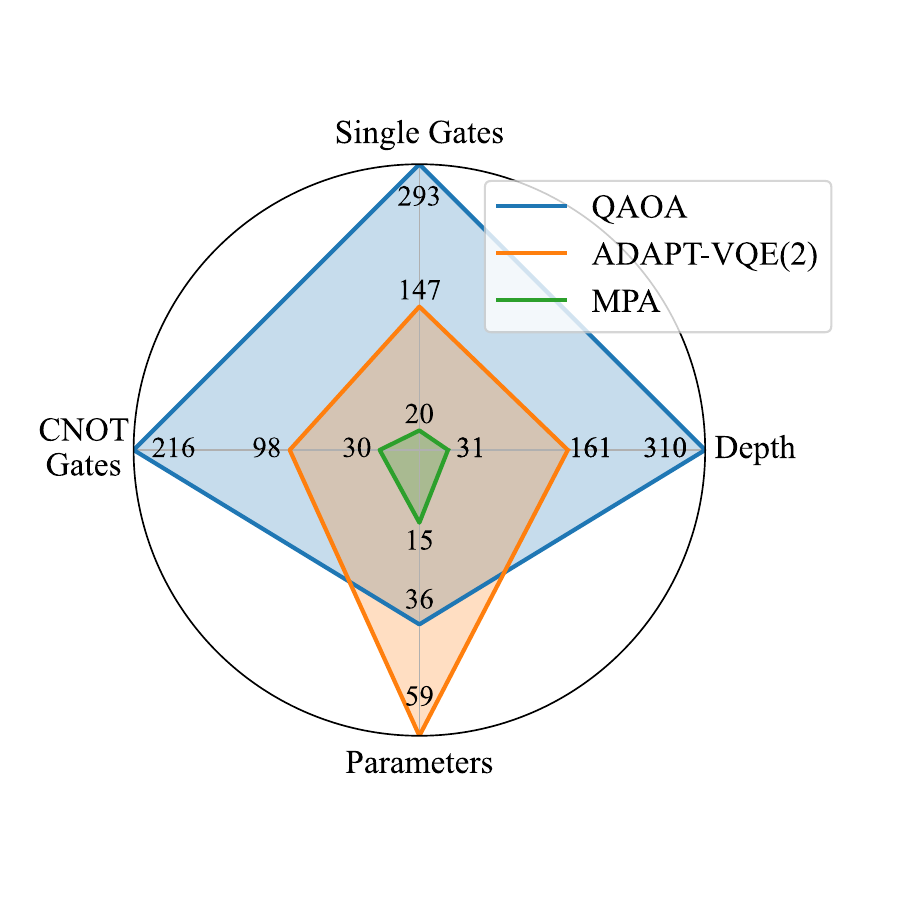}
    \caption{Comparison of the circuit complexity between QAOA, ADAPT-VQE(2), and the MPA circuit introduced in this work for the frustrated 5-site house Ising model with $a=0.1$.
}
\label{fig.9}
\end{figure}

We analyzed the RCE and state fidelity relative to the exact results for $a = 0.1$, where the first excitation gap is $1.5\times10^{-4}$. The results are presented in Table \ref{tab.2} and visualized in Fig.~\ref{fig.9}. As shown in the results, for a small value of $a$, the QAOA method requires a large number of layers to achieve accurate energy and wave function results. However, even with a depth of 310, the QAOA circuit does not reach a fidelity of 1.000 and exhibits a higher energy error compared to the MPA method. For ADAPT-VQE, a gradient threshold of $10^{-5}$ is required to achieve a fidelity of 1.000. However, with a gradient threshold of $10^{-4}$, it is possible to obtain accurate energy results, with an energy error lower than that of the proposed method, but at the cost of poor state fidelity. Using very small gradient thresholds increases the circuit depth and also raises the number of measurements required for the operators in the pool. This example illustrates that the MPA circuit can achieve accurate results with the highest fidelity while significantly reducing circuit complexity, requiring only one-tenth the depth of the QAOA circuit and one-fifth that of the ADAPT-VQE (see Fig.~\ref{fig.9}).
\begin{table*}[htbp]
    \centering
    \caption{Comparison of energy error, \%RCE, and $F$ between QAOA, ADAPT-VQE, and
the MPA circuit introduced in this work for the frustrated 5-site house
Ising model with $a=0.1$. ADAPT-VQE(1) and ADAPT-VQE(2) correspond to
different choices of the gradient convergence threshold in the adaptive
operator-selection procedure.}
    \label{tab.2}
    \renewcommand{\arraystretch}{1.3}
    \begin{tabular}{ccccc}
        \toprule
        Method & Energy Error & \%RCE & Fidelity & Gradient Threshold  \\
        \midrule
        QAOA & 0.01495 & -721.4 & 0.954 & - \\
        ADAPT-VQE(1) & 0.00008 & 95.6 & 0.500 & $10^{-4}$  (4 iterations) \\
        ADAPT-VQE(2) & 0.00000 & 100 & 1.000 & $10^{-5}$ (5 iterations) \\
        MPA & 0.00006 & 96.4 & 1.000 & - \\
        \bottomrule
    \end{tabular}
\end{table*}

\subsection{7 and 10-site frustrated Ising model}
\label{subsec:5}
To further evaluate the capability of the MPA circuit, the lattice size was increased to 7 and 10 sites to determine the number of $s$ and $t$ layers required to achieve accurate results for \%RCE and $F$ within $a$. The notation $\textrm{MPA}_{m}$ denotes an ansatz composed of $m$ successive layers, where each layer consists of an $s-t$ pair acting collectively on the full qubit register (See Fig.~\ref{fig.s-t}).

\begin{figure}[htbp]
\Qcircuit @C=1.8em @R=0.4em { &&&&\mbox{\ \ \ \ \ \ \ \ \ \ \ \ $m$  }&&&&\\
&&&&&&&&\\
&&&&&&&&\\
& &&&\mbox{\  2 \ \ \ \ \ \ \ \ \ }  &&&&&\\ 
&&&&&&&&\\
&&&&&&&&\\
&&&&&&&&\\
&&&\mbox{   1 \ \ \ \ \ \ \ \ \ } &&&&&&\\
&&&&&&&&\\
&&&&&& \cdots&&&&\\
& \qw & \multigate{3}{s} & \multigate{3}{t} &  \multigate{3}{s} & \multigate{3}{t}& \qw & \multigate{3}{s} & \multigate{3}{t}& \qwa \\ 
& \qw &  \ghost{s} &  \ghost{t} & \ghost{s} &   \ghost{t}   & \qw  & \ghost{s} &   \ghost{t}& \qwa \\ 
&\vdots & \nghost{s}& \nghost{t} & \nghost{s} &  \nghost{t} &\cdots & & \nghost{s} &  \nghost{t}&  \\ 
& \qw & \ghost{s}   &  \ghost{t} & \ghost{s} &     \ghost{t}  &  \qw &  \ghost{s} &     \ghost{t}& \qwa 
\gategroup{11}{3}{12}{4}{.7em}{^)}
\gategroup{7}{3}{8}{6}{.7em}{^)}
\gategroup{3}{3}{4}{9}{.7em}{^)}
}
\caption{Schematic representation of the \(\mathrm{MPA}_m\) ansatz, where \(m\) denotes the number of \(s\!-\!t\) pairs used.}
    \label{fig.s-t}
\end{figure}

Fig.~\ref{fig.10} illustrates a representative Max-Cut configuration among the four distinct solutions obtained for the 7-site hexagon graph using the QAOA algorithm. The corresponding results are shown in Fig.~\ref{fig.11}, indicating that an accurate solution corresponds to $\textrm{MPA}_3$.

\begin{figure}[htbp]
    \centering
   \begin{tikzpicture}[scale=0.2, every node/.style={circle, draw, minimum size=0.2cm, font=\bfseries}]

\node[fill=bluegreen!70] (n0) at (-10.5, 0) {0};
\node[fill=red!70]  (n1) at (-7,-7) {1};
\node[fill=bluegreen!70] (n2) at (0,-7) {2};
\node[fill=red!70]  (n3) at (3.5,0) {3};
\node[fill=bluegreen!70] (n4) at (0,7) {4};
\node[fill=red!70]  (n5) at (-7,7) {5};
\node[fill=red!70]  (n6) at (-3.5,0) {6};

\draw (n0) -- (n1);
\draw (n0) -- (n5);
\draw (n0) -- (n6);
\draw (n1) -- (n2);
\draw (n1) -- (n6);
\draw (n2) -- (n3);
\draw (n2) -- (n6);
\draw (n3) -- (n4);
\draw (n3) -- (n6);
\draw (n4) -- (n5);
\draw (n4) -- (n6);
\draw (n5) -- (n6);

\draw[dashed , ultra thick, black]
  (-5,10) .. controls (-1,4)..(7,2)

.. controls (4,-5)..(-1,-4)

.. controls (-8,-12)..(-10,-5)
.. controls (-6,0)..(-12,6);
\end{tikzpicture}
    \caption{One possible Max-Cut for the 7-site hexagon graph.}
    \label{fig.10}
\end{figure}

\begin{figure}[htbp]
\centering
\subfloat[]{
    \includegraphics[width=0.4\textwidth]{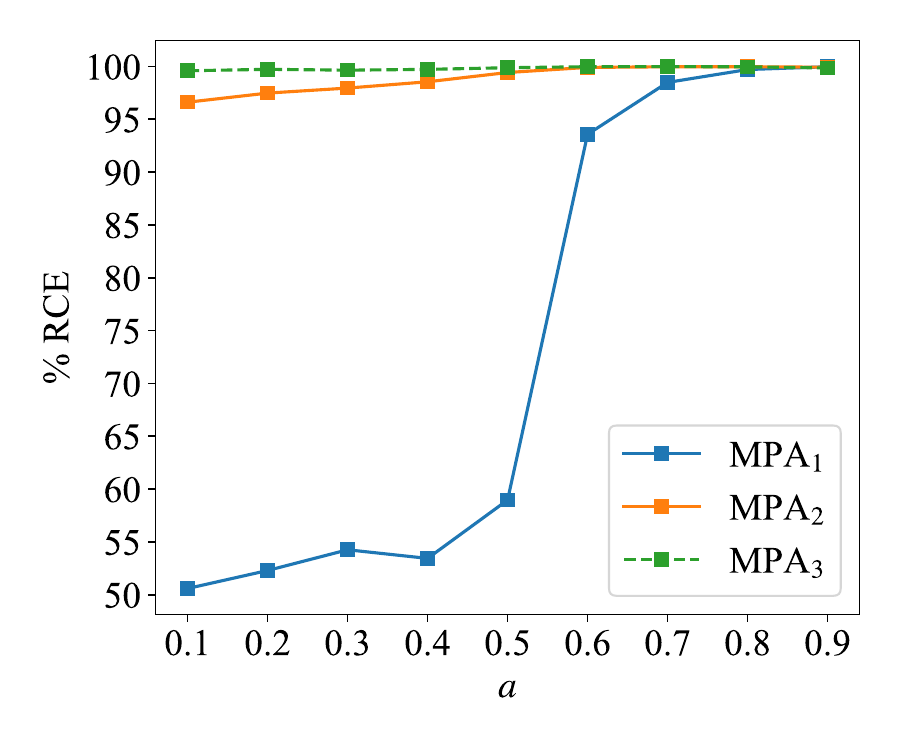}
    \centering
\label{fig.11(a)}
}
\hspace{0.06\textwidth}
\subfloat[]{
    \includegraphics[width=0.4\textwidth]{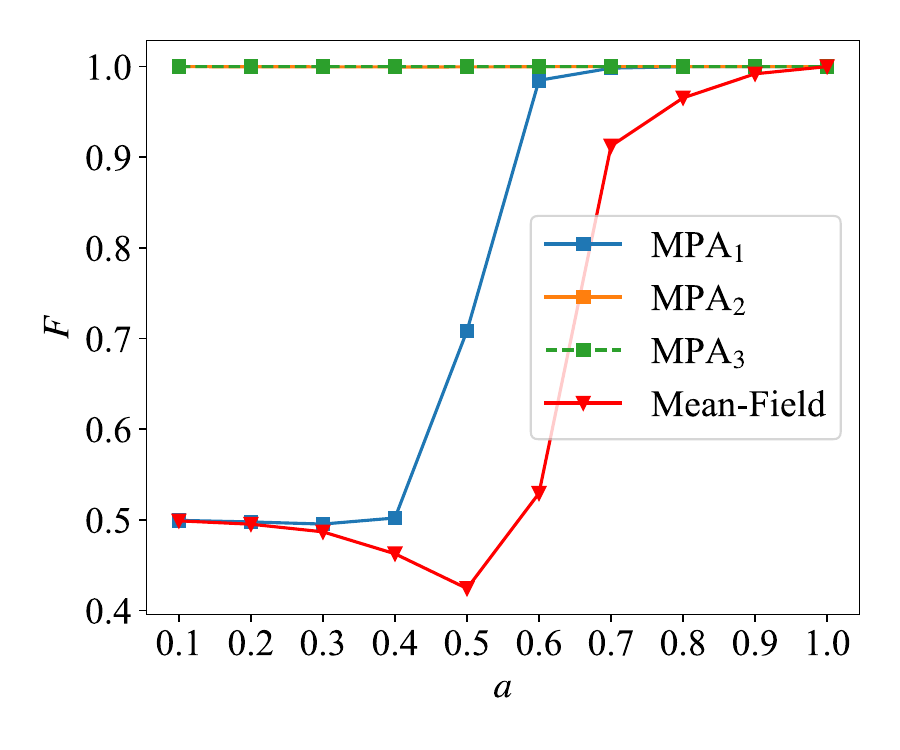}
    \centering
\label{fig.11(b)}
}
\caption{Comparison of the recovered correlation energy for different $m$ shown in (a), and the state fidelity presented in (b) for the 7-site hexagon frustrated Ising model. The mean-field curve corresponds to a product-state ansatz constructed
using only single-qubit \(R_y\) rotations (see Sec.~III.B).}
 
\label{fig.11}
\end{figure}
For the 10-site ribbon lattice, one possible Max-Cut configuration was obtained using the QAOA algorithm from a set of six distinct solutions, together with an alternative configuration (Figs.~\ref{fig.12} and~\ref{fig.13}). The simulation was carried out according to the configuration shown in Fig.~\ref{fig.12}, while a comparison between the two Max-Cut solutions, including the results for the alternative configuration, is provided in the Appendix \ref{a2}. Interestingly, $m=3$ once again provides a result with high state fidelity and accurate energy, as shown in Fig.~\ref{fig.14}. 

\begin{figure}[htbp]  
    \centering
    \begin{tikzpicture}[scale=0.2, every node/.style={circle, draw, minimum size=0.2cm, font=\bfseries}]

\node[fill=bluegreen!70] (n0) at (-10.5, 0) {0};
\node[fill=red!70]  (n1) at (-7,-7) {1};
\node[fill=bluegreen!70] (n2) at (0,-7) {2};
\node[fill=red!70]  (n3) at (7,-7) {3};
\node[fill=bluegreen!70] (n4) at (10,0) {4};
\node[fill=red!70]  (n5) at (7,7) {5};
\node[fill=bluegreen!70] (n6) at (0,7) {6};
\node[fill=red!70]  (n7) at (-7,7) {7};
\node[fill=red!70]  (n8) at (-3.5,0) {8};
\node[fill=red!70]  (n9) at (3.5,0) {9};

\draw (n0) -- (n1);
\draw (n0) -- (n7);
\draw (n0) -- (n8);
\draw (n1) -- (n2);
\draw (n1) -- (n8);
\draw (n2) -- (n3);
\draw (n2) -- (n8);
\draw (n2) -- (n9);
\draw (n3) -- (n4);
\draw (n3) -- (n9);
\draw (n4) -- (n5);
\draw (n4) -- (n9);
\draw (n5) -- (n6);
\draw (n5) -- (n9);
\draw (n6) -- (n7);
\draw (n6) -- (n8);
\draw (n6) -- (n9);
\draw (n7) -- (n8);
\draw (n8) -- (n9);

\draw[dashed , ultra thick, black]
  (-5,10) .. controls (0,2)..(6,9)
.. controls (9,10)..(9.5,4)
.. controls (6,0)..(9,-4)
.. controls (10,-8)..(7,-10)   
.. controls (5,-9)..(3,-6)   
.. controls (1,-3.5)..(-1,-4)  
.. controls (-4,-8)..(-7,-10)  
.. controls (-11,-9)..(-8,-4)
.. controls (-6,0)..(-12,6)

   ;

\end{tikzpicture}
    \caption{One of the six possible Max-Cut solutions for the 10-site ribbon graph}
    \label{fig.12}
\end{figure}

\begin{figure}[htbp]  
    \centering
    \begin{tikzpicture}[scale=0.2, every node/.style={circle, draw, minimum size=0.2cm, font=\bfseries}]

\node[fill=red!70] (n0) at (-10.5, 0) {0};
\node[fill=bluegreen!70]  (n1) at (-7,-7) {1};
\node[fill=red!70] (n2) at (0,-7) {2};
\node[fill=bluegreen!70]  (n3) at (7,-7) {3};
\node[fill=red!70] (n4) at (10,0) {4};
\node[fill=bluegreen!70]  (n5) at (7,7) {5};
\node[fill=red!70] (n6) at (0,7) {6};
\node[fill=bluegreen!70]  (n7) at (-7,7) {7};
\node[fill=bluegreen!70]  (n8) at (-3.5,0) {8};
\node[fill=red!70]  (n9) at (3.5,0) {9};

\draw (n0) -- (n1);
\draw (n0) -- (n7);
\draw (n0) -- (n8);
\draw (n1) -- (n2);
\draw (n1) -- (n8);
\draw (n2) -- (n3);
\draw (n2) -- (n8);
\draw (n2) -- (n9);
\draw (n3) -- (n4);
\draw (n3) -- (n9);
\draw (n4) -- (n5);
\draw (n4) -- (n9);
\draw (n5) -- (n6);
\draw (n5) -- (n9);
\draw (n6) -- (n7);
\draw (n6) -- (n8);
\draw (n6) -- (n9);
\draw (n7) -- (n8);
\draw (n8) -- (n9);

\draw[dashed , ultra thick, black]
(-12,5)

.. controls (-6,0).. (-10,-8)

.. controls (-8,-12).. (-3,-7)

.. controls (0,2).. (-3,8)

.. controls (1,12).. (5,3)

.. controls (5,3).. (12,2)

.. controls (13,-5).. (4,-4)

.. controls (2,-8).. (5,-11)
   ;

\end{tikzpicture}
    \caption{Alternative Max-Cut configuration for the 10-site ribbon graph.}
    \label{fig.13}
\end{figure}

\begin{figure}[htbp]
\centering
    \subfloat[]{
    \includegraphics[width=0.4\textwidth]{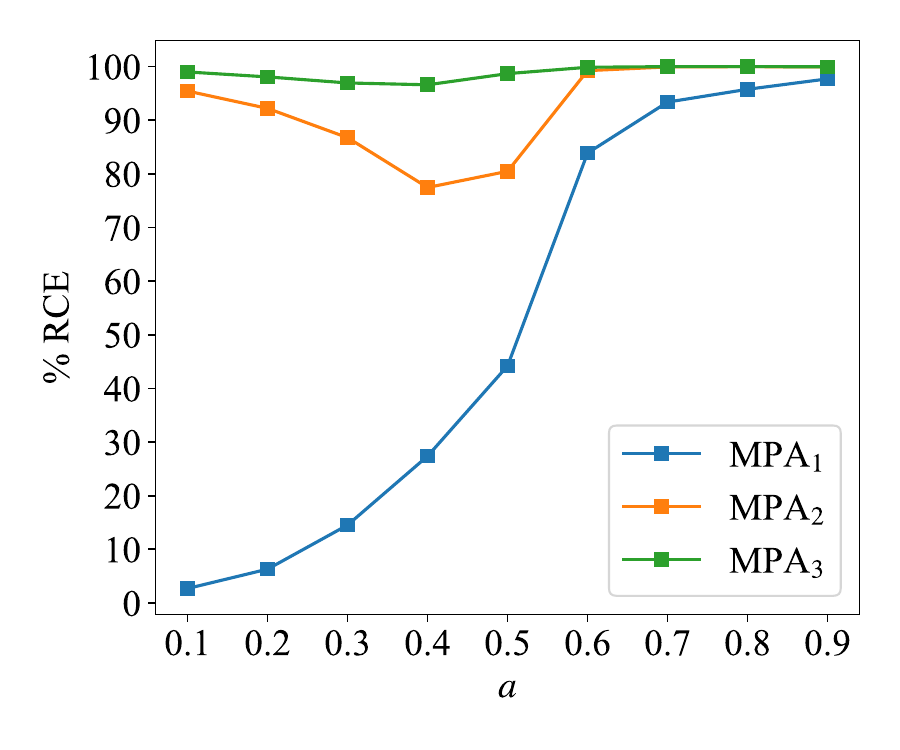}
   \label{fig.14(a)}
}
 \hspace{0.06\textwidth}
    \subfloat[]{
    \includegraphics[width=0.4\textwidth]{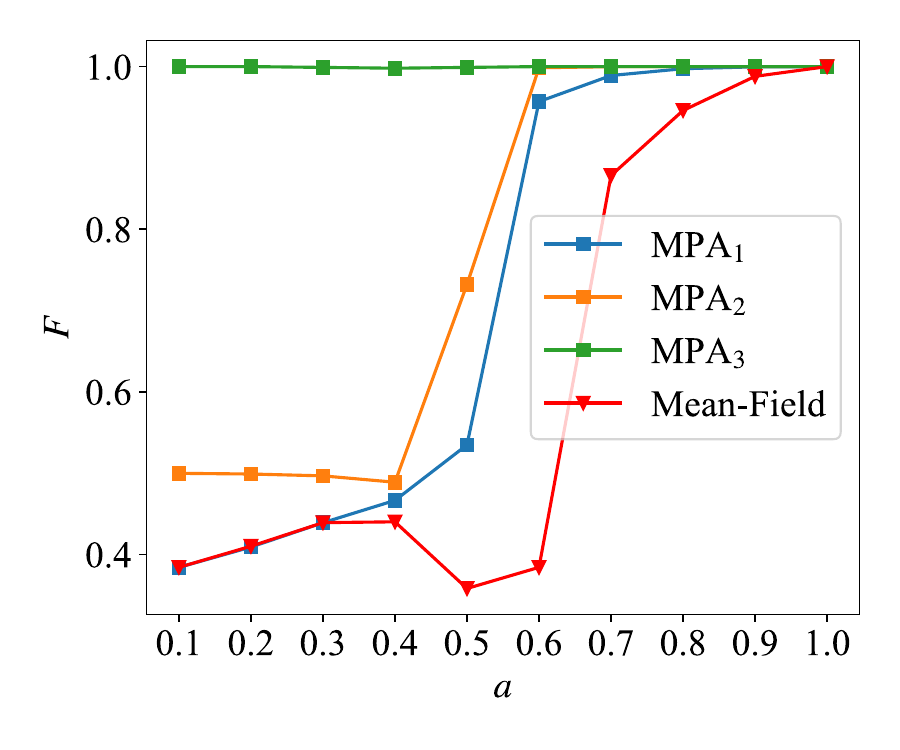}
\label{fig.14(b)}
}
\caption{Comparison of the recovered correlation energy for different $m$ shown in (a), and the state fidelity presented in (b) for the 10-site ribbon frustrated Ising model. }
\label{fig.14}
\end{figure}

In Table \ref{tab.3}, \%RCE and $F$ compared with QAOA and ADAPT-VQE for $a=0.1$, where the first excitation gap is $2.861\times 10^{-8}$. For the QAOA method, 100 layers of problem and mixer Hamiltonians were utilized \cite{farhi2014quantum}. However, as evident from the table, despite this substantial number of repetitions, the state fidelity remains very poor, and the energy error is significant. Due to the extremely small first excitation gap, achieving accurate results would require repeating the layers an excessively large number of times, leading to an exponential increase in circuit depth and the number of CNOT gates. On the other hand, the ADAPT-VQE method provides accurate energy estimates. However, with a gradient threshold of $10^{-4}$, the state fidelity is not satisfactory, as this configuration results in a circuit with substantial depth and an excessive number of parameters. Consequently, obtaining an output state with high fidelity using ADAPT-VQE becomes challenging without significantly reducing the gradient threshold. Lowering the threshold further would produce a circuit that is both extremely deep and heavily parameterized, making it impractical. Table \ref{tab.4} illustrates a comparison of circuit complexity scaling from a 5-site to a 10-site system for the MPA circuit, QAOA, and ADAPT-VQE with $a=0.1$. As observed, the circuit depth for the QAOA method increases significantly, yet the fidelity remains very poor. For the ADAPT-VQE approach, the scaling is approximately linear, similar to our method; however, the fidelity for the 10-site system is poor, and both the circuit depth and the number of parameters are significantly higher. Notably, the circuit depth for the 10-site system using our method is still lower than the depth required for the 5-site system with the ADAPT-VQE method.

To further assess the robustness of the proposed circuit beyond uniform
interactions, we also investigate a frustrated 10-site ribbon lattice with
non-uniform (random) couplings. This additional benchmark, presented in
Appendix~\ref{a3}, demonstrates that the circuit retains its
performance under interaction inhomogeneity.

\begin{table*}[!htbp]
    \centering
    \caption{Comparison of energy error,\%RCE and $F$ between QAOA, ADAPT-VQE, and the MPA circuit introduced in this work for the frustrated 10-site ribbon Ising model with $a=0.1$.}
    \label{tab.3}
    \renewcommand{\arraystretch}{1.3}
    \begin{tabular}{ccccc}
        \toprule
        Method & Energy Error & \%RCE & Fidelity & Gradient Threshold  \\
        \midrule
        QAOA & 1.74323 & -4107.5 & 0.067 & - \\
        ADAPT-VQE& 0.00027 & 99.3 & 0.500 & $10^{-4}$  (3 iterations) \\
        MPA & 0.00042 & 99.0 & 1.000 & - \\
        \bottomrule
    \end{tabular}
\end{table*}
\begin{table}[!htbp]
\centering
\caption{Comparison of circuit complexity scaling for different methods across different system sizes.}
\label{tab.4}
\begin{tabular}{ccccc}
\toprule
\textrm{Sites} & \textrm{Method} & \textrm{Depth} & \textrm{CNOT} & \textrm{Parameters} \\
\midrule
\multirow{3}{*}{5}  & QAOA        & 310  & 216  & 36  \\
                    & ADAPT-VQE(2)   & 161  & 98   & 59  \\
                    & MPA   & 31   & 30   & 15  \\
\midrule
\multirow{3}{*}{10} & QAOA        & 4701 & 3800 & 200 \\
                    & ADAPT-VQE   & 551  & 342  & 194 \\
                    & MPA   & 121  & 168  & 84  \\
\bottomrule
\end{tabular}
\end{table}

In the following section, to evaluate the scalability of the proposed method for larger lattices and to verify whether $m=3$ continues to provide a high-fidelity output state, we extend the lattice size to 20 sites.
\subsection{19 \& 20-site frustrated Ising model}
\label{subsec:6}
We applied the same procedure to simulate a lattice with 20 sites, corresponding to a Hilbert space dimension of $\mathrm{dim}\,\mathcal{H} \simeq 10^6$. One possible Max-Cut configuration, selected from forty solutions obtained using QAOA as the initial step in the circuit design, is shown in Fig.~\ref{fig.15}. Due to the lack of symmetry in this lattice, compared to the previous lattices, it serves as a useful test for evaluating the capability of the MPA circuit. We started from $a=1$ and employed adiabatic optimization. For $m=3$ and $m=4$, the results are shown in Fig.\ref{fig.16}, and Table \ref{tab.5} presents the circuit complexity.
\begin{figure*}[!htbp]        
    \centering
    \begin{tikzpicture}[scale=0.2, every node/.style={circle, draw, minimum size=0.8cm, font=\bfseries}]


\node[fill=red!70] (0) at (-11, 0) {0};
\node[fill=bluegreen!70] (1) at (-7,-7) {1};
\node[fill=red!70] (2) at (0,-7) {2};
\node[fill=bluegreen!70] (3) at (7,-7) {3};
\node[fill=red!70] (4) at (14,-7) {4};
\node[fill=bluegreen!70] (5) at (21,-7) {5};
\node[fill=red!70] (6) at (28,-7) {6};
\node[fill=bluegreen!70] (7) at (31,0) {7};
\node[fill=bluegreen!70] (8) at (35,7) {8};
\node[fill=red!70] (9) at (28,7) {9};
\node[fill=bluegreen!70] (10) at (21,7) {10};
\node[fill=red!70] (11) at (14,7) {11};
\node[fill=bluegreen!70] (12) at (7,7) {12};
\node[fill=red!70] (13) at (0,7) {13};
\node[fill=bluegreen!70] (14) at (-7,7) {14};
\node[fill=bluegreen!70] (15) at (-4,0) {15};
\node[fill=red!70] (16) at (3,0) {16};
\node[fill=bluegreen!70] (17) at (10,0) {17};
\node[fill=red!70] (18) at (17,0) {18};
\node[fill=red!70] (19) at (24,0) {19};

\draw (0) -- (1);
\draw (0) -- (15);
\draw (0) -- (14);
\draw (1) -- (2);
\draw (1) -- (15);
\draw (2) -- (15);
\draw (2) -- (16);
\draw (2) -- (3);
\draw (3) -- (4);
\draw (3) -- (16);
\draw (3) -- (17);
\draw (4) -- (5);
\draw (4) -- (17);
\draw (4) -- (18);
\draw (5) -- (19);
\draw (5) -- (6);
\draw (5) -- (18);
\draw (6) -- (7);
\draw (6) -- (19);
\draw (7) -- (8);
\draw (7) -- (9);
\draw (7) -- (19);
\draw (8) -- (9);
\draw (9) -- (10);
\draw (9) -- (19);
\draw (10) -- (19);
\draw (10) -- (18);
\draw (10) -- (11);
\draw (11) -- (12);
\draw (11) -- (17);
\draw (11) -- (18);
\draw (12) -- (13);
\draw (12) -- (17);
\draw (12) -- (16);
\draw (13) -- (14);
\draw (13) -- (15);
\draw (13) -- (16);
\draw (14) -- (15);
\draw (15) -- (16);
\draw (16) -- (17);
\draw (17) -- (18);
\draw (18) -- (19);

\end{tikzpicture}
    \caption{One possible Max-Cut for the 20-site ribbon graph.}
    \label{fig.15}
\end{figure*}

\begin{figure}[htbp]
\centering
    \subfloat[]{
    \includegraphics[width=0.4\textwidth]{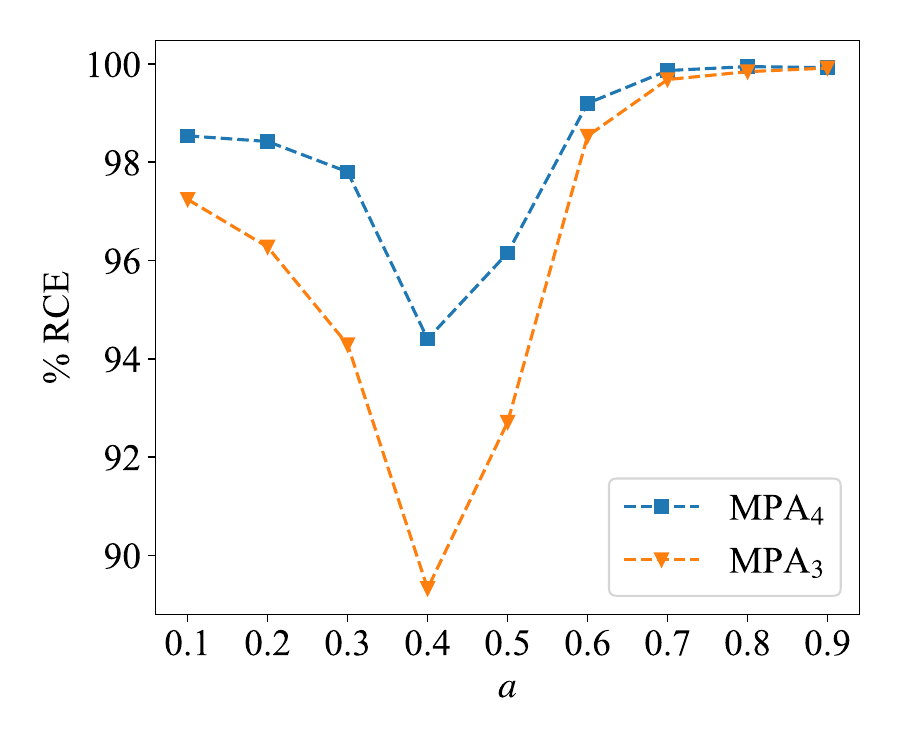}
    \label{fig.16(a)}
}
\hspace{0.06\textwidth}
    \subfloat[]{
    \includegraphics[width=0.4\textwidth]{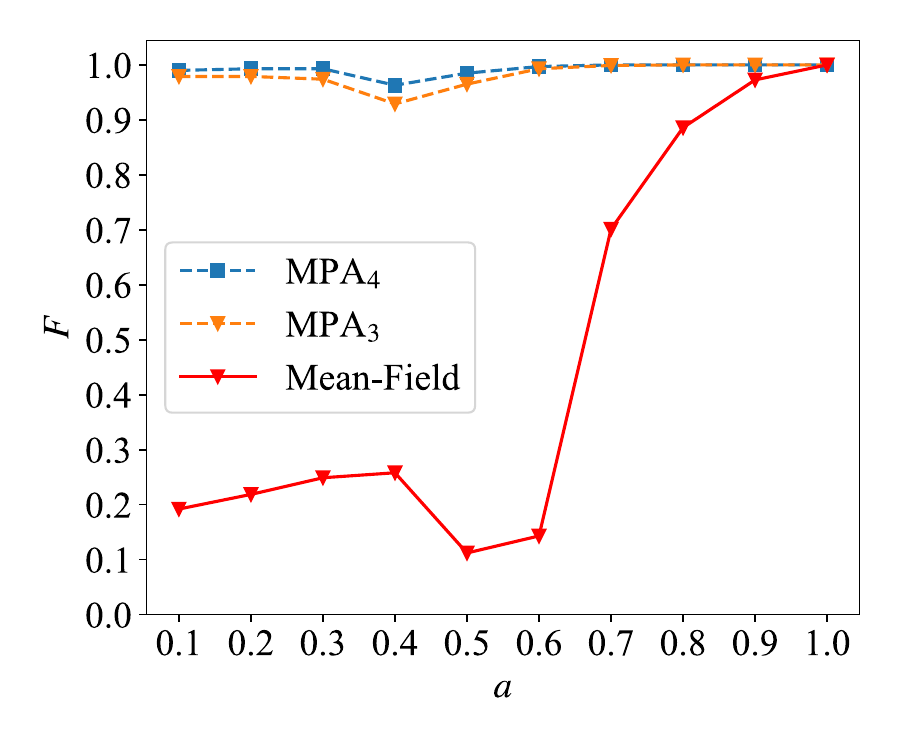}
    \centering
    \label{fig.16(b)}
    }
\caption{\%RCE for different $m$ shown in (a), and $F$ presented in (b) for the 20-site ribbon frustrated Ising model.}
\label{fig.16}
\end{figure}

The results indicate that $m=3$ is adequate to reproduce the target state with high fidelity across nearly the entire interval of $a$. However, the energy error can be further reduced by increasing $m$ to 4. 
Despite the fact that the first excitation gap decreases exponentially with increasing lattice size in the weak-field regime, as shown in Fig.~\ref{fig.17}, the MPA circuit demonstrates remarkable robustness by consistently maintaining high fidelity.

\begin{figure*}[!htbp]
    \centering
    \includegraphics[width=0.8\textwidth]{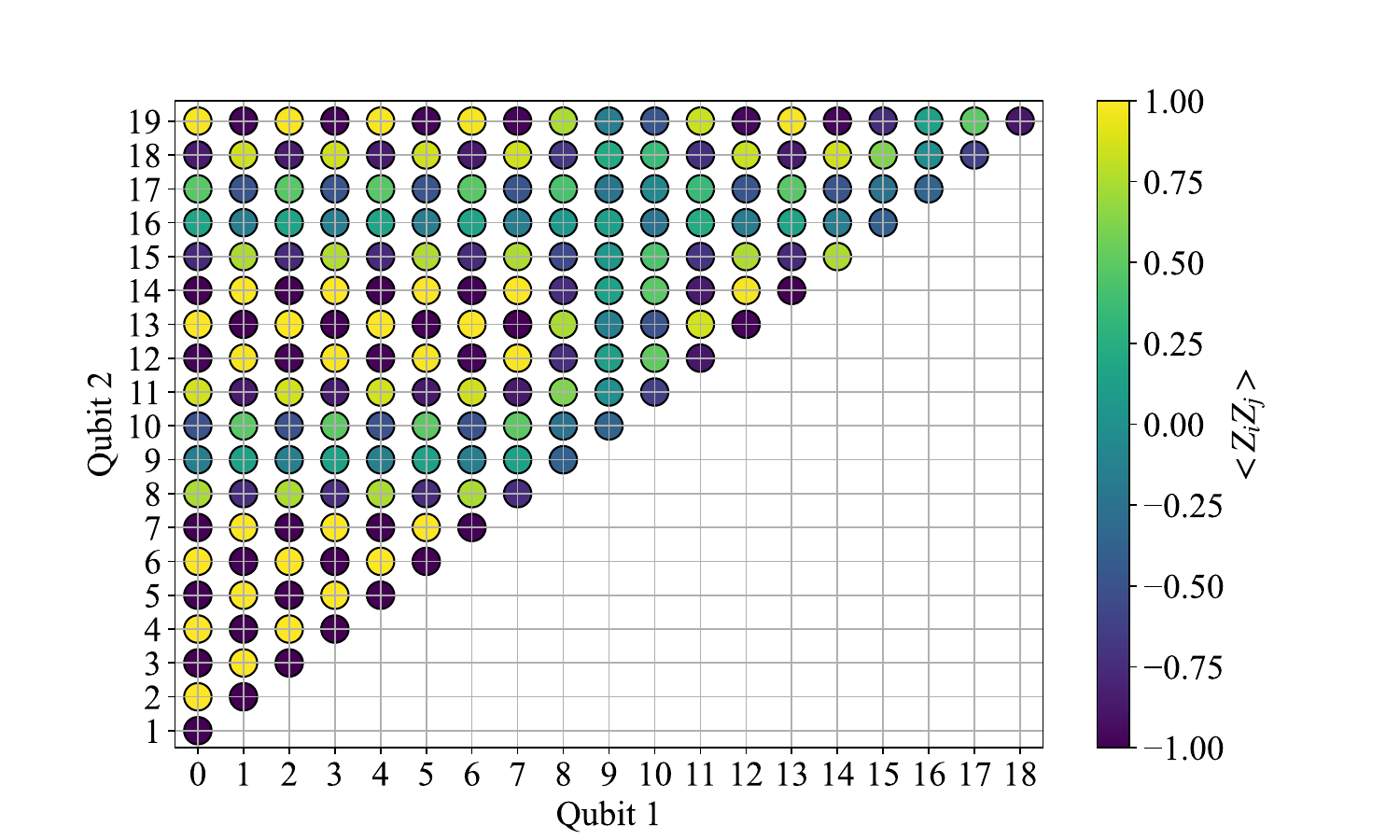}
    \caption{The correlation function obtained from the exact results for the 20-site ribbon system.
}
    \label{fig.18}
\end{figure*}

\begin{figure*}[!htbp]
    \centering
    \includegraphics[width=0.8\textwidth]{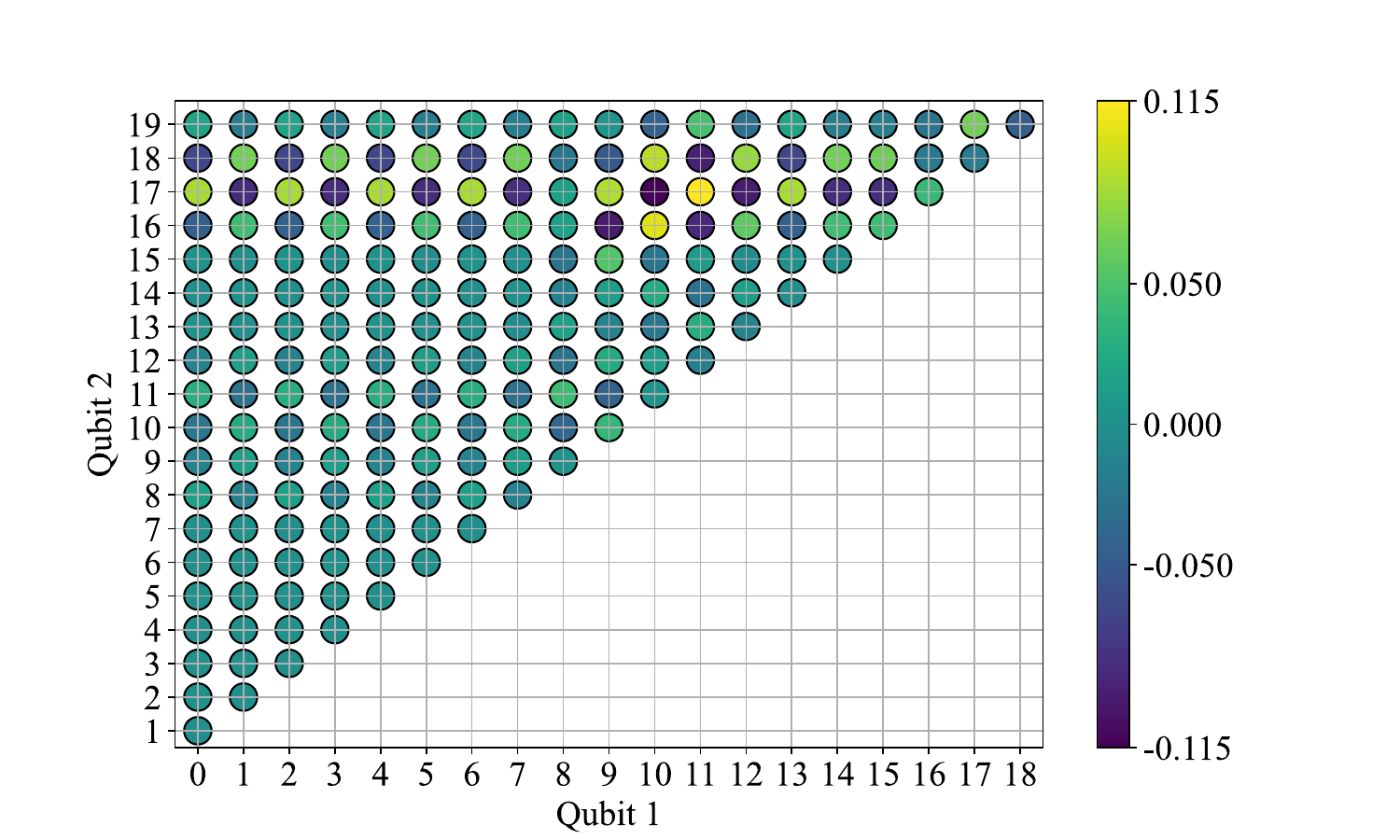}
    \caption{The difference between the correlation function for the exact results and the simulated results for the 20-site ribbon system.
\label{fig.19}}
\end{figure*}

\begin{figure}[htbp]
\centering
    \subfloat[]{%
    \includegraphics[width=0.45\textwidth]{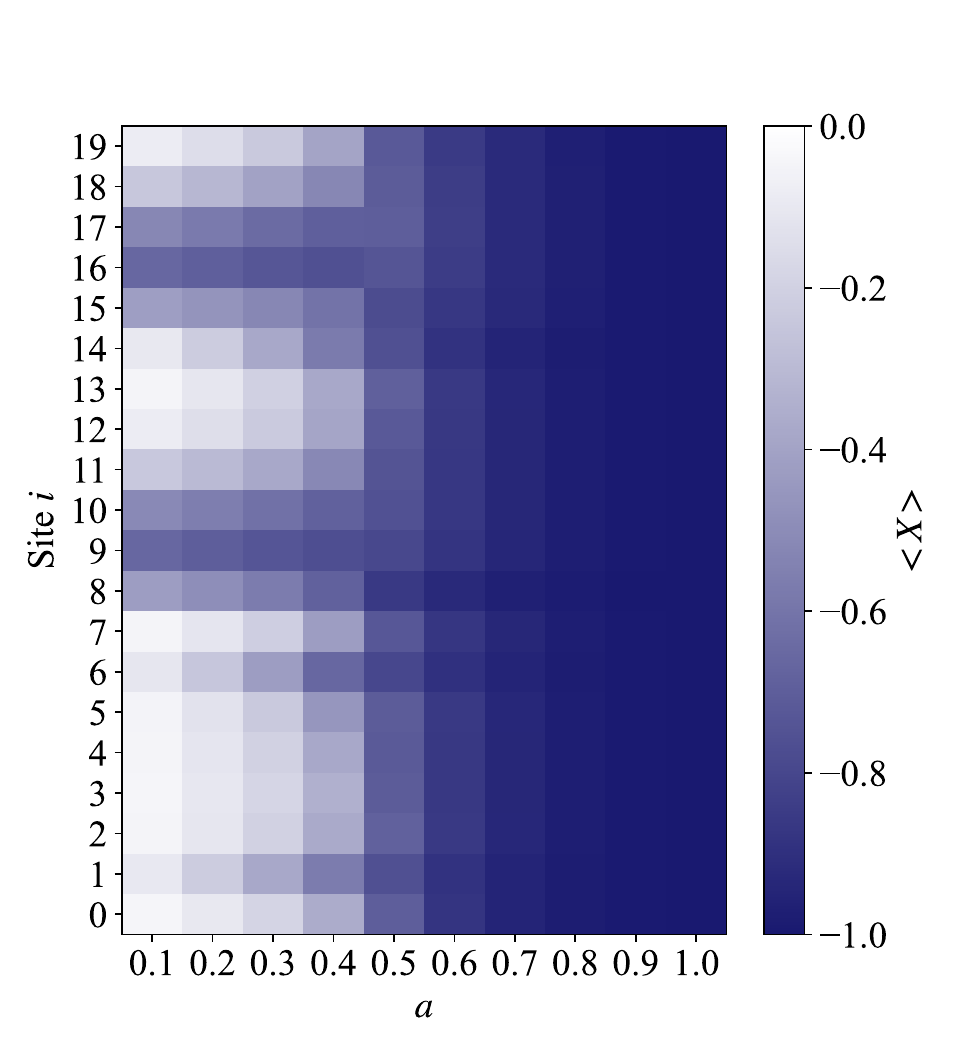}
    \label{fig.20(a)}
}
\hspace{0.06\textwidth}
    \subfloat[]{%
    \includegraphics[width=0.45\textwidth]{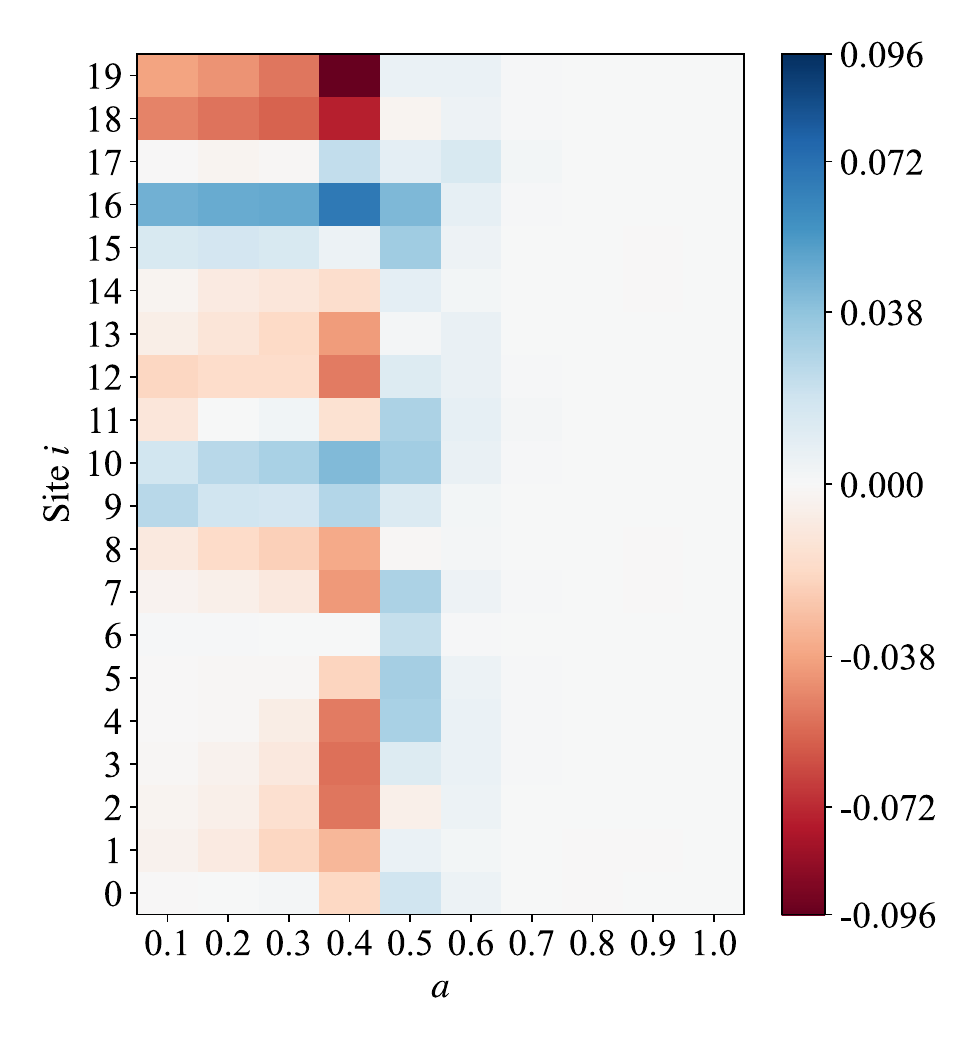}
    \label{fig.20(b)}
}
\caption{The magnetization obtained from the exact results for the 20-site ribbon system is shown in (a), and the difference between the magnetization obtained from the exact results and the simulated results is shown in (b).
}
\label{fig.20}
\end{figure}

We continue evaluating the performance of the circuit with $m=3$ to analyze the correlation function $\langle Z_iZ_j\rangle$ and the magnetization $\langle X_i\rangle$, comparing them with the exact results. Calculations were performed for $a = 0.1$, which corresponds to the most challenging regime, by evaluating all possible interactions across all sites on the obtained output state and comparing them with the exact results. Fig.~\ref{fig.18} presents the correlation function for the exact results, while Fig.~\ref{fig.19} illustrates the differences with the simulation results. The results indicate that the error in the correlation function is small on average; however, larger deviations of up to $|\pm 0.115|$ are observed for interactions involving sites 16, 17, and 18. Consistently, by analyzing the magnetization, we observe that despite acceptable agreement with the exact data for small values of $a$, sites 16, 18, and 19 exhibit larger errors compared to other sites, with a maximum absolute error of 0.096 (see Fig.~\ref{fig.20}).

\begin{figure}[htbp]    
    \centering
    \begin{tikzpicture}[scale=1.5, every node/.style={circle, draw, minimum size=0.8cm, font=\bfseries}]

\node[fill=red!70] (0) at (0,4) {0};
\node[fill=bluegreen!70] (1) at (1,4) {1};
\node[fill=red!70] (2) at (2,4) {2};

\node[fill=bluegreen!70] (3) at (-0.5,3) {3};
\node[fill=red!70] (4) at (0.5,3) {4};
\node[fill=bluegreen!70] (5) at (1.5,3) {5};
\node[fill=red!70] (6) at (2.5,3) {6};

\node[fill=red!70] (7) at (-1,2) {7};
\node[fill=bluegreen!70] (8) at (0,2) {8};
\node[fill=bluegreen!70] (9) at (1,2) {9};
\node[fill=red!70] (10) at (2,2) {10};
\node[fill=bluegreen!70] (11) at (3,2) {11};

\node[fill=bluegreen!70] (12) at (-0.5,1) {12};
\node[fill=red!70] (13) at (0.5,1) {13};
\node[fill=red!70] (14) at (1.5,1) {14};
\node[fill=bluegreen!70] (15) at (2.5,1) {15};

\node[fill=red!70] (16) at (0,0) {16};
\node[fill=bluegreen!70] (17) at (1,0) {17};
\node[fill=red!70] (18) at (2,0) {18};

\draw (0) -- (1);
\draw (0) -- (4);
\draw (0) -- (3);
\draw (1) -- (5);
\draw (1) -- (4);
\draw (1) -- (2);
\draw (2) -- (5);
\draw (2) -- (6);
\draw (3) -- (4);
\draw (3) -- (7);
\draw (3) -- (8);
\draw (4) -- (5);
\draw (4) -- (8);
\draw (4) -- (9);
\draw (5) -- (6);
\draw (5) -- (9);
\draw (5) -- (10);
\draw (6) -- (10);
\draw (6) -- (11);
\draw (7) -- (8);
\draw (7) -- (12);
\draw (8) -- (12);
\draw (8) -- (13);
\draw (8) -- (9);
\draw (9) -- (10);
\draw (9) -- (13);
\draw (9) -- (14);
\draw (10) -- (11);
\draw (10) -- (14);
\draw (10) -- (15);
\draw (11) -- (15);
\draw (12) -- (13);
\draw (12) -- (16);
\draw (13) -- (14);
\draw (13) -- (16);
\draw (13) -- (17);
\draw (14) -- (15);
\draw (14) -- (17);
\draw (14) -- (18);
\draw (15) -- (18);
\draw (16) -- (17);
\draw (17) -- (18);

\end{tikzpicture}
    \caption{One of the forty possible Max-Cut configurations for the 19-site hexagon graph.}
    \label{fig.21}
\end{figure}

\begin{figure*}[!htbp]    
    \centering
    \includegraphics[width=0.8\textwidth]{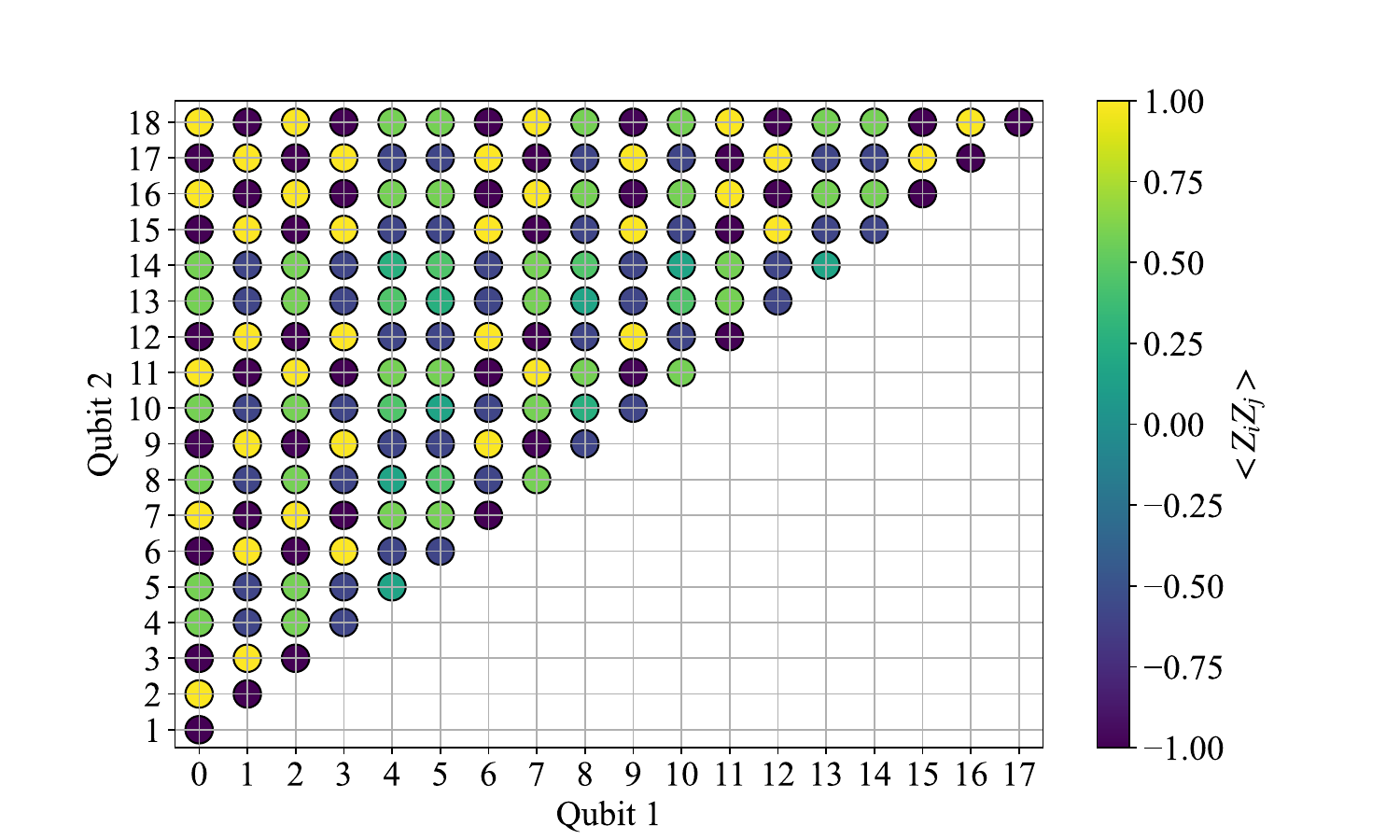}
    \caption{The correlation function obtained from the exact results for the 19-site  hexagon system.}
    \label{fig.22}
\end{figure*}

\begin{figure*}[!htbp]   
    \centering
    \includegraphics[width=0.8\textwidth]{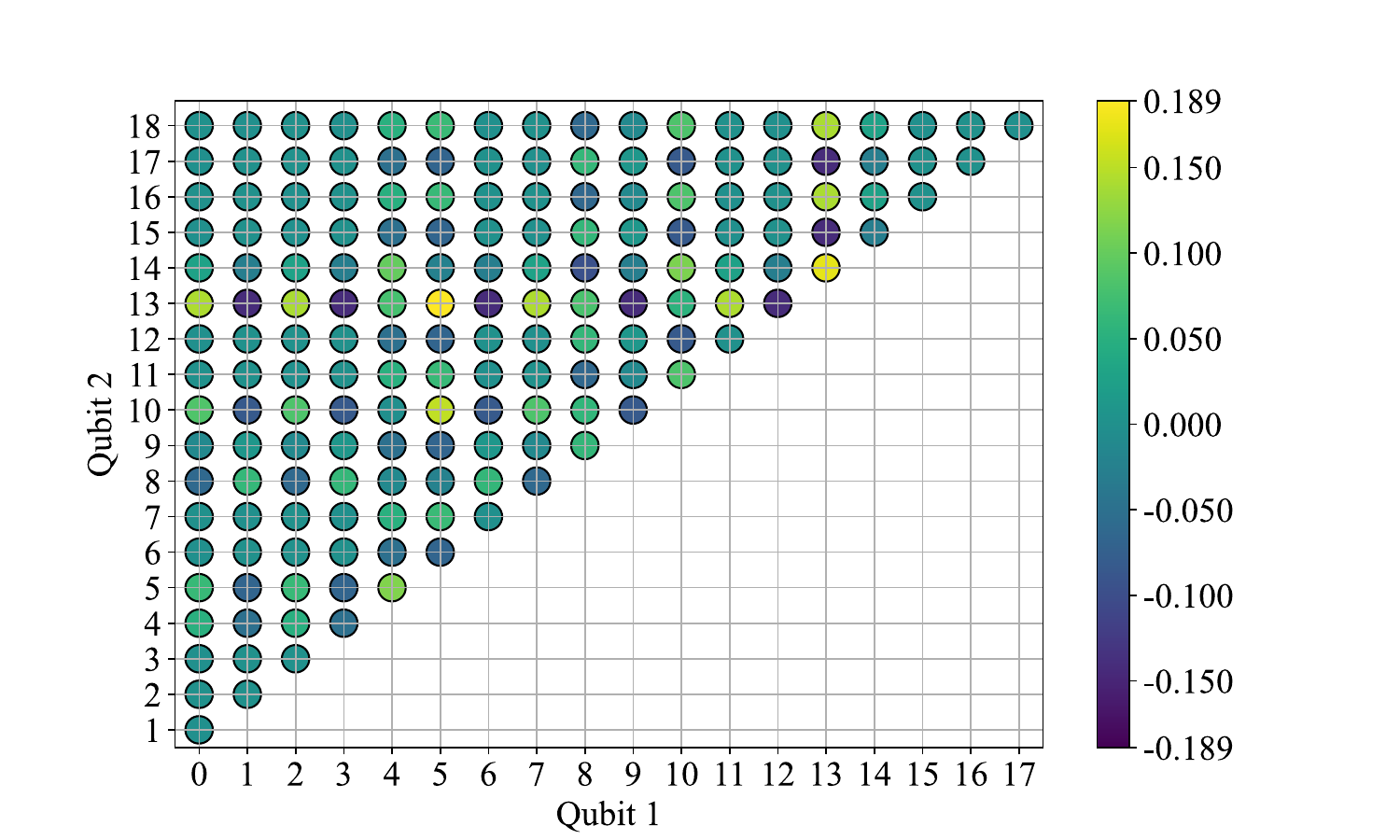}
    \caption{The difference between the correlation function for the exact results and the simulated results for the 19-site  hexagon system.}
    \label{fig.23}
\end{figure*}

From the position of sites 16, 18 and 19, we hypothesize that interactions in the bulk are more difficult to capture, and so we conducted an additional simulation for a 19-site hexagon lattice (Fig.~\ref{fig.21}), which has a configuration that is slightly bulkier than the 20-site ribbon lattice. Figs.\ref{fig.22} and \ref{fig.23} present the results for the exact values and the differences between the exact and simulated results for $a=0.1$, respectively. The output state fidelity is 0.972, with an energy error of $1.242\times10^{-2}$. The circuit complexity corresponding to the 19-site lattice is summarized in Table~\ref{tab.5}. The results demonstrate that the proposed method successfully produces a state with high fidelity. Additionally, the correlation function derived from the simulation shows good agreement for most interactions. Notably, for site 9, which is located at the center of the graph and is farthest from the edges, the error in the correlation function involving this site is relatively low. 
\begin{table}[htbp]
    \centering
    \caption{Summary of the CNOT gate count and the number of parameters in the MPA circuit for different system sizes, evaluated for \( m = 3 \).}
    \label{tab.5}
    \renewcommand{\arraystretch}{1.2}
    \begin{tabular}{cccc}
        \toprule
      \textrm{Name} &  \textrm{Sites (\(n\))} & \textrm{CNOT} & \textrm{Parameters} \\
        \midrule
      House & 5  & 30  & 15  \\
      Hexagon & 7  & 108 & 54  \\
      Ribbon & 10 & 168 & 84  \\
      Hexagon & 19 & 360 & 180 \\
      Ribbon & 20 & 360 & 180 \\
        \bottomrule
    \end{tabular}
\end{table}

\begin{figure}[htbp]
\centering
    \subfloat[]{%
    \includegraphics[width=0.45\textwidth]{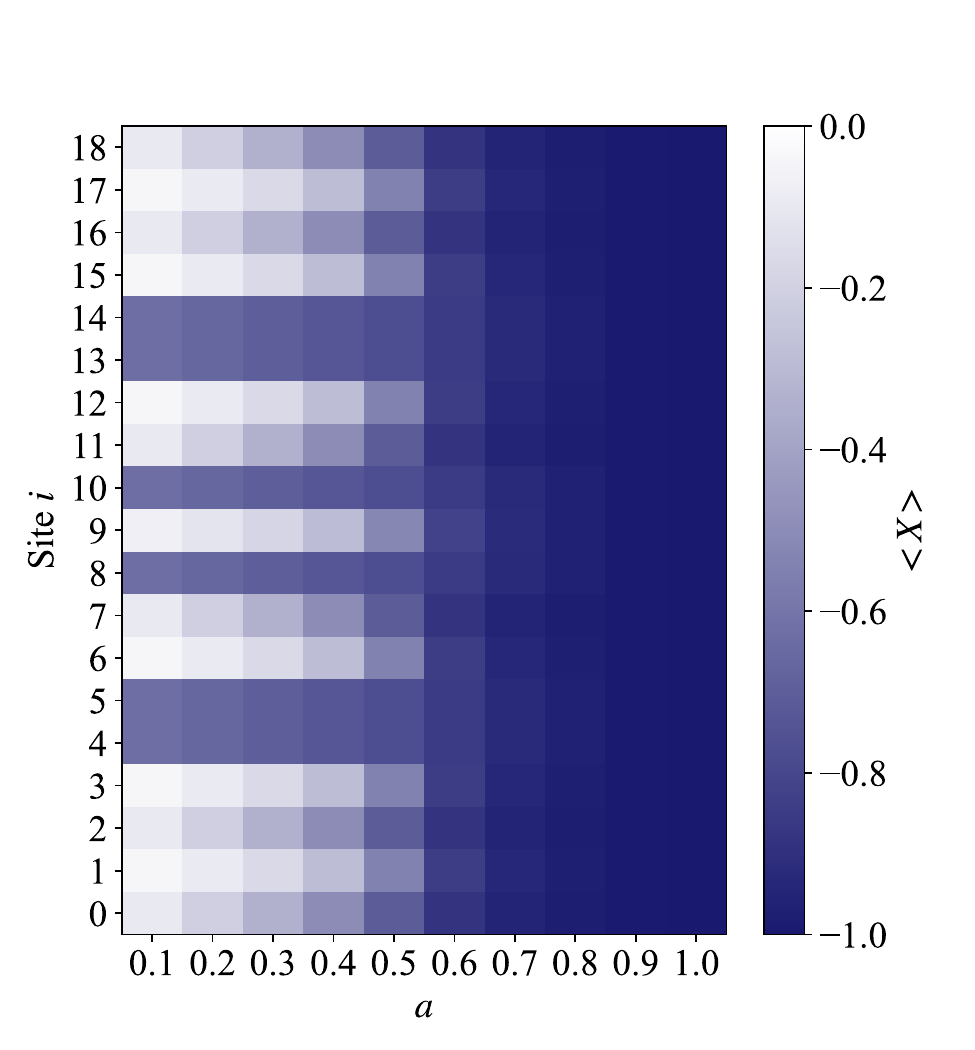}
\label{fig.24(a)}
}
\hspace{0.06\textwidth}
    \subfloat[]{%
    \includegraphics[width=0.45\textwidth]{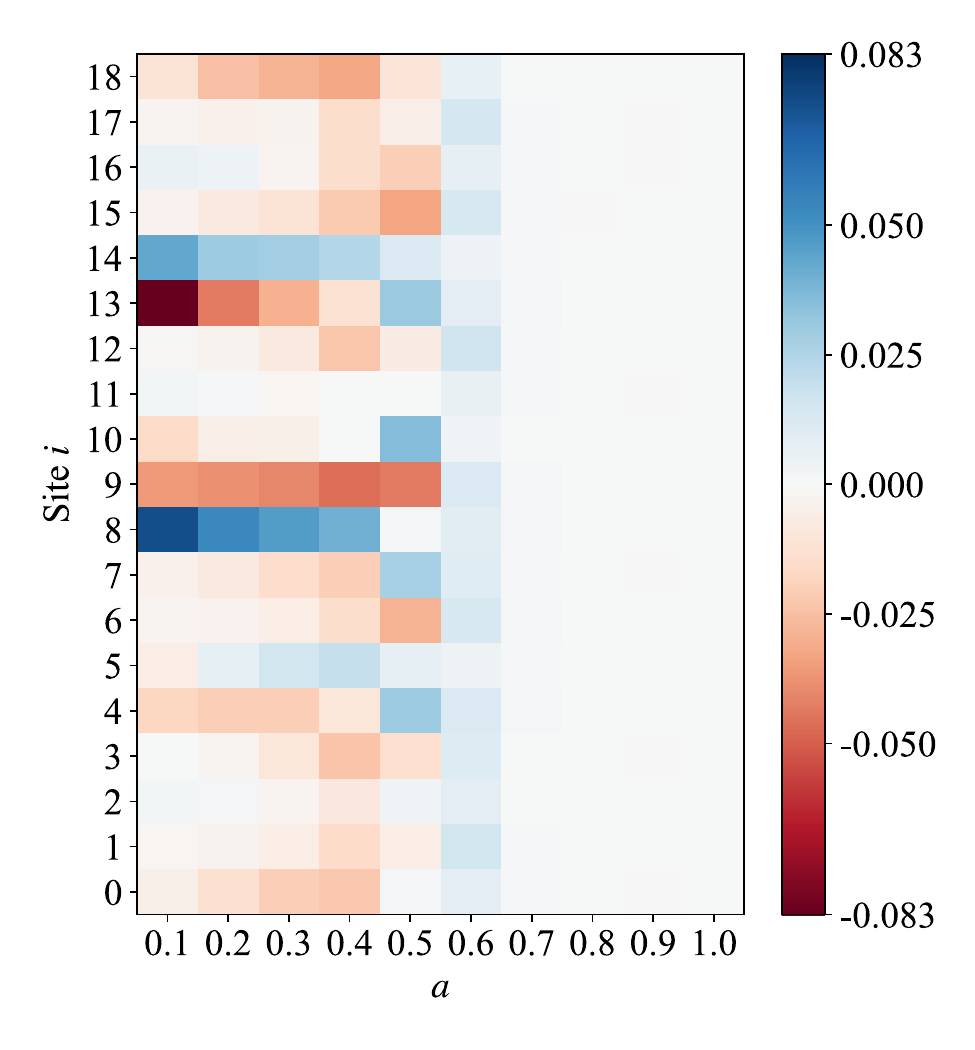}
\label{fig.24(b)}
}
\caption{The magnetization obtained from the exact results for the 19-site  hexagon system is shown in (a), and the difference between the magnetization obtained from the exact results and the simulated results is shown in (b). }
\label{fig.24}
\end{figure}

Moreover, the magnetization was calculated and compared to the exact result, as shown in Fig.~\ref{fig.24}. The values obtained from the circuit approximate the exact results well, with the highest absolute error being 0.083. As illustrated in Fig.~\ref{fig.24(b)}, although sites 5 and 13 are expected to behave identically due to the lattice symmetry, the MPA circuit breaks this symmetry, leading to a larger error at site 13 than 5.

\section{Conclusion}
\label{sec:3}
We introduced a 2-qc block to design a variational quantum circuit for Ising spin glasses using the VQE framework. Graph partitioning of the underlying frustrated lattice based on the Max-Cut problem is utilized to minimize circuit depth. The method has been tested on frustrated lattices with up to 20 sites. Notably, when scaling the circuit from a 10-site to a 20-site lattice, the number of CNOT gates, circuit depth, and required parameters increases by a factor of 2, despite the exponential growth in the number of possible configurations, exceeding one million for the 20-site lattice. In the RVB picture, the number of possible pairings for a 20-site lattice exceeds 650 million, compared to only 945 pairings for a 10-site lattice. This demonstrates the exponential growth in circuit depth required to accommodate all possible pairings, highlighting the importance of methods that can reduce the search space by identifying significant pairings. By employing graph partitioning based on Max-Cut, the circuit complexity is significantly reduced. 

In general, finding a Max-Cut is an NP-hard problem. The most effective classical polynomial-time approximation algorithm is that of Goemans and Williamson~\cite{goemans1995improved}, which employs semidefinite programming combined with randomized rounding and achieves an approximation ratio of approximately 88\% of the maximum cut. For quantum algorithms, single-layer QAOA ($p=1$) achieves an approximation ratio of about 69\% for Max-Cut on 3-regular graphs~\cite{farhi2014quantum}. Nevertheless, for planar graphs, classical polynomial-time methods exist that reformulate the Max-Cut problem as a maximum-weight matching problem solvable using polynomial-time algorithms~\cite{hadlock1975finding}.

In terms of gate scaling, for a planar graph $G(V,E)$ with $V$ sites and $|E|$ edges, let $k$ denote the number of edges cut in the maximum cut of $G$, which is bounded by $|E|/2 \le k \le |E|$. A single layer of the MPA therefore requires $4k$ CNOT gates and $2k$ variational parameters, while a single QAOA layer uses $2|E|$ CNOT gates and two parameters. Consequently, the per-layer CNOT count of MPA ranges between $1\times$ and $2\times$ that of QAOA. Although QAOA is more parameter-efficient per layer, achieving comparable performance typically requires multiple layers.

Based on the results, \( m = 3 \) provides acceptable accuracy compared to exact solutions, meaning that the circuit complexity also remains polynomial. It should be mentioned that the seemingly constant $m=3$ was not tested on larger systems than 20 spins, because of computational limitations on the exact benchmark results, which require exact diagonalization in an exponentially scaling Hilbert space. However, if a constant or polynomially scaling $m$ with system size can be confirmed, the MPA circuit design might open a door to polynomial scaling simulation of RVB spin glasses on quantum devices.

Finally, the 2-qc block exhibits a group structure, as this operation is closed for any input in the Hilbert space. Additionally, the commutation properties with the total-spin flip operator, such as the Hamiltonian of the problem, can provide potential insights by examining the corresponding Lie algebraic structure and the dynamical Lie algebra of the building block. This aspect will be addressed in future work.

\begin{acknowledgments}
SDB and SEG acknowledge the Canada Research Chair program, the CFI, NSERC Discovery Grant program, and NBIF for financial support.
\end{acknowledgments}

\section*{Data Availability}
A Jupyter Notebook of the simulations can be found at Ref.~\cite{qunb_qcchem}.

\clearpage

\appendix

\section{QAOA and ADAPT-VQE}
\label{a1}
In our QAOA simulations of the frustrated Ising model in Eq.~(\ref{eq:TFIM}), the cost-Hamiltonian evolution was implemented using a first-order Trotter decomposition~\cite{Nielsen,QiskitEvolvedOperatorAnsatz}. The mixer Hamiltonian was chosen as
\( H_M = \sum_{i=0}^{n-1} Y_i \),
since the Ising interaction terms act in the \(Z\) direction while the transverse magnetic field is applied along \(X\).
This choice follows the general QAOA requirement that the mixer Hamiltonian should not commute with the problem Hamiltonian, thereby enabling nontrivial exploration of the Hilbert space~\cite{farhi2014quantum,bako2025prog}.
We applied 18 alternating layers of the problem and mixer Hamiltonians for the 5-site house geometry and 100 layers for the 10-site ribbon lattice.

For the ADAPT-VQE method, an operator pool was constructed using single-qubit and two-qubit gates to implement single and double excitations, representing spin flips at lattice sites. We have the following candidate operators for constructing the pool:
\begin{align*}
\text{Single-Qubit Operators:} & \quad \{\sigma_{i}^{\alpha}\}.\\
\text{Two-Qubit Operators:} & \quad \{\sigma_{i}^{\alpha}\sigma_{j}^{\beta}\}.
\end{align*}
The two-qubit operators included in the pool are derived from neighbor-neighbor interactions based on the graph structure. The gradient of the energy with respect to each operator is initially evaluated for the two-qubit operators. The selected operators are then implemented in a quantum circuit, which is initialized in the vacuum state, with their parameters optimized. In the subsequent step, the gradient for a single operator is measured and added to the existing circuit, followed by re-optimizing all parameters within the circuit. This iterative process continues until the gradient threshold and convergence tolerance are achieved. 
\section{Comparative Analysis of Max-Cut Configurations (10-Site ribbon Lattice)}
\label{a2}
\begin{figure}[htbp]
    \centering
    \subfloat[]{%
        \includegraphics[width=0.4\textwidth]{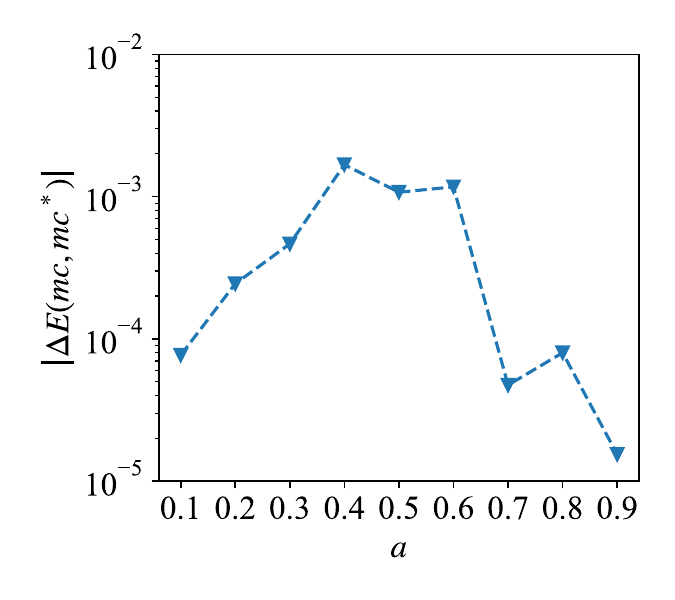}
        \label{fig.A1a}
}
    \hspace{0.05\textwidth}
    \subfloat[]{%
        \includegraphics[width=0.4\textwidth]{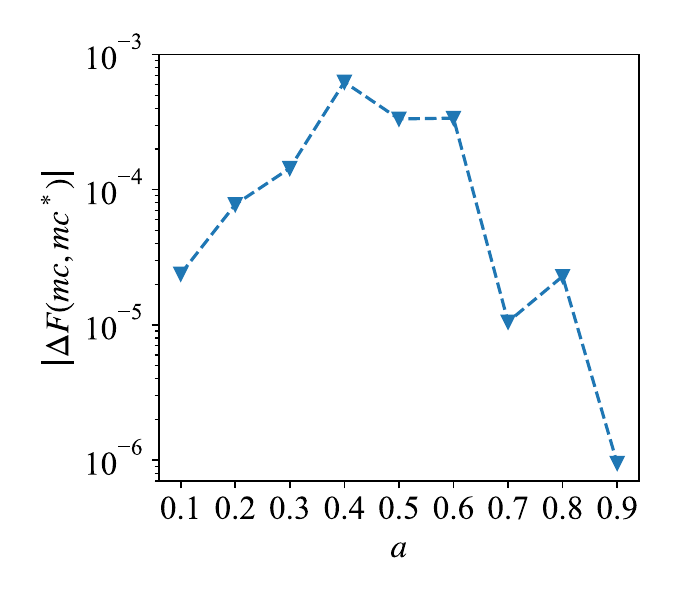}
        \label{fig.A1b}
}
    \caption{Comparison of two Max-Cut configurations in the 10-site ribbon lattice.}
    \label{fig.A1}
\end{figure}

Here, we present a comparison between two distinct Max-Cut configurations in terms of energy error and fidelity for the 10-site ribbon lattice. We denote the configuration shown in Fig.~\ref{fig.12} as $mc$ and the one in Fig.~\ref{fig.13} as $mc^*$. The energies obtained from the corresponding circuits for $mc$ and $mc^*$ are compared in Fig.~\ref{fig.A1a}. The absolute energy error, $|\Delta E(mc, mc^*)|$, demonstrates strong agreement between the two results, with the error remaining below $10^{-4}$ for $a = 0.1$, which represents the most challenging regime. 

Furthermore, the difference in fidelity between the two output states, $\Delta F(mc, mc^*)$, with respect to the true ground state, has been computed and is shown in Fig.~\ref{fig.A1b}. The results indicate that the fidelity difference remains below $10^{-3}$ across the entire range of $a$, and notably below $10^{-4}$ for $a = 0.1$, suggesting that the two distinct Max-Cut configurations yield nearly identical output states.

\section{Random interactions for the 10-site ribbon lattice}
\label{a3}

As an additional benchmark, we investigate the performance of the MPA method in the
presence of non-uniform antiferromagnetic interactions. Specifically, we consider a
representative instance in which the coupling strengths are randomly drawn from a
uniform distribution on the interval $[0,1]$, as listed in Table~\ref{tab:random}.
The first excitation gap is evaluated to demonstrate that this lattice configuration
also exhibits an exponentially small gap.

The Max-Cut partitioning for this lattice with non-uniform edge weights is illustrated
in Fig.~\ref{fig.A2}, where the resulting source and sink layers of the graph are
identified.

The corresponding energy error and state fidelity are shown in
Figs.~\ref{fig.A3a} and~\ref{fig.A3b}, respectively. These results indicate that the
proposed circuit accurately reproduces the target ground state, achieving a fidelity
of $1.000$.

\begin{figure}[htbp]  
    \centering
    \begin{tikzpicture}[scale=0.2, every node/.style={circle, draw, minimum size=0.2cm, font=\bfseries}]

\node[fill=bluegreen!70] (n0) at (-10.5, 0) {0};
\node[fill=red!70]  (n1) at (-7,-7) {1};
\node[fill=red!70] (n2) at (0,-7) {2};
\node[fill=bluegreen!70]  (n3) at (7,-7) {3};
\node[fill=bluegreen!70] (n4) at (10,0) {4};
\node[fill=bluegreen!70]  (n5) at (7,7) {5};
\node[fill=red!70] (n6) at (0,7) {6};
\node[fill=red!70]  (n7) at (-7,7) {7};
\node[fill=bluegreen!70]  (n8) at (-3.5,0) {8};
\node[fill=red!70]  (n9) at (3.5,0) {9};

\draw (n0) -- (n1);
\draw (n0) -- (n7);
\draw (n0) -- (n8);
\draw (n1) -- (n2);
\draw (n1) -- (n8);
\draw (n2) -- (n3);
\draw (n2) -- (n8);
\draw (n2) -- (n9);
\draw (n3) -- (n4);
\draw (n3) -- (n9);
\draw (n4) -- (n5);
\draw (n4) -- (n9);
\draw (n5) -- (n6);
\draw (n5) -- (n9);
\draw (n6) -- (n7);
\draw (n6) -- (n8);
\draw (n6) -- (n9);
\draw (n7) -- (n8);
\draw (n8) -- (n9);

\draw[dashed , ultra thick, black]
(-12,-5)

.. controls (3,-1).. (-5,4)

.. controls (-15,9).. (1,10)

.. controls (4,7).. (6,4)

.. controls (7,3).. (6,-4)

.. controls (5,-5).. (1,-10)

   ;

\end{tikzpicture}
    \caption{One of the two possible Max-Cut solutions for the weighted 10-site ribbon graph.}
    \label{fig.A2}
\end{figure}

\begin{table}[h!]
\centering
\caption{
Random nearest--neighbour couplings $J_{ij}$ sampled independently from 
the uniform distribution on $[0,1]$, i.e., $J_{ij} \sim \mathcal{U}(0,1)$.
}
\begin{tabular}{c c c}
\hline
$i$ & $j$ & $J_{ij}$ \\
\hline
0 & 1 & 0.78742 \\
0 & 7 & 0.89209 \\
0 & 8 & 0.01508 \\
1 & 2 & 0.83367 \\
1 & 8 & 0.54790 \\
2 & 3 & 0.97345 \\
2 & 9 & 0.23683 \\
2 & 8 & 0.64692 \\
3 & 4 & 0.06503 \\
3 & 9 & 0.55558 \\
4 & 9 & 0.40236 \\
4 & 5 & 0.26597 \\
5 & 9 & 0.98668 \\
5 & 6 & 0.39732 \\
8 & 6 & 0.32846 \\
6 & 9 & 0.12516 \\
6 & 7 & 0.11146 \\
7 & 8 & 0.98457 \\
9 & 8 & 0.57894 \\
\hline
\end{tabular}
\label{tab:random}
\end{table}

\begin{figure}[htbp]    
   \centering
    \includegraphics[width=0.4\textwidth]{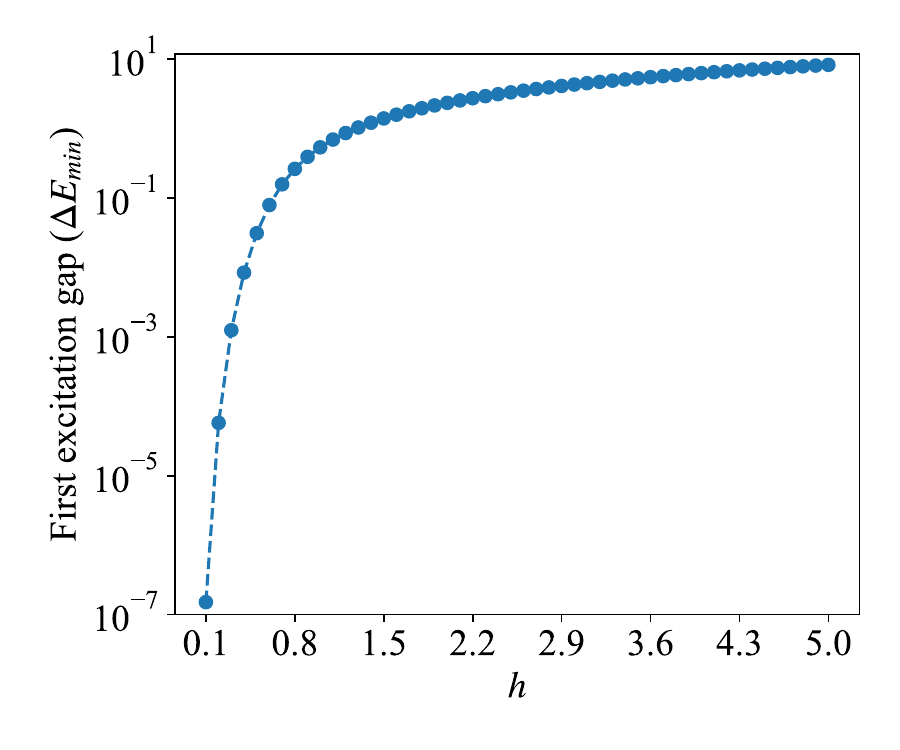}
    \caption{First excitation gap of the 10-site ribbon lattice with non-uniform interactions, evaluated across different magnetic-field strengths \( h \).}
    \label{fig.27}
\end{figure}

\begin{figure}[htbp]
    \centering
    \subfloat[]{%
        \includegraphics[width=0.4\textwidth]{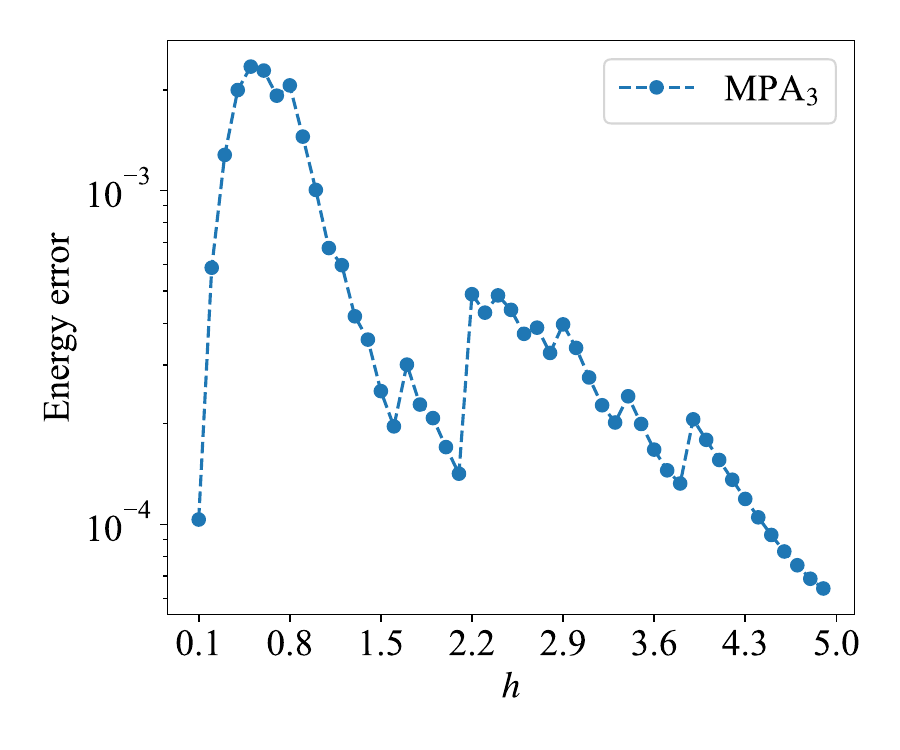}
        \label{fig.A3a}
}
    \hspace{0.05\textwidth}
    \subfloat[]{%
        \includegraphics[width=0.4\textwidth]{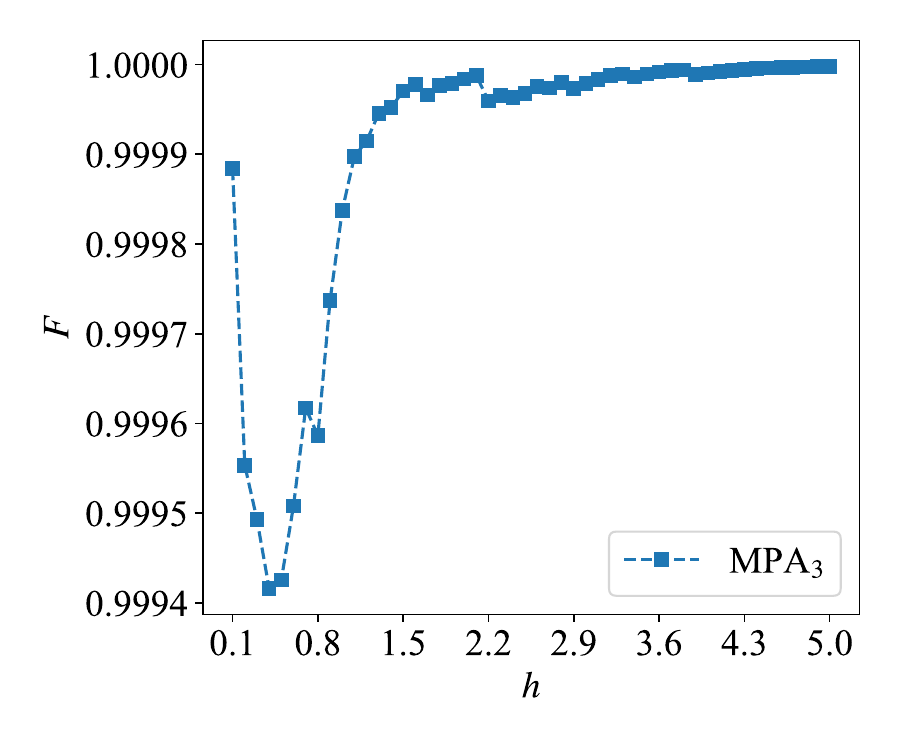}
        \label{fig.A3b}
}
    \caption{Energy error in (a) and fidelity results (b) for $\textrm{MPA}_{3}$ the 10-site ribbon lattice with non-uniform interactions.}
    \label{fig.A3}
\end{figure}

\section{Hardware-Realistic Noise Simulations}
\label{app:noise}

In this appendix, we assess the robustness of the proposed method under
finite-shot sampling and hardware-realistic noise.
We consider three representative geometries: the 5-site house graph and
the 10-site and 20-site ribbon graphs.

For each system, we compare energy estimates obtained from
(i) infinite-shot simulations,
(ii) finite-shot ideal simulations with 4000 measurement shots, and
(iii) finite-shot noisy simulations with 4000 measurement shots.
The infinite-shot results serve as a reference, while the finite-shot
simulations quantify the effects of statistical sampling and
device-level imperfections.

All noisy simulations were performed locally using Qiskit Aer, with a
noise model constructed from the IBM Marrakesh backend~\cite{ibm_marrakesh_backend}.
The backend configuration was used to extract the native basis gates,
qubit coupling map, and calibrated gate and readout noise parameters,
thereby enabling hardware-realistic simulations.

To isolate the effects of shot noise and hardware-induced errors,
finite-shot ideal and noisy energies were evaluated at the optimal
variational parameters obtained from the corresponding infinite-shot
simulations.

\begin{figure}[htbp]    
   \centering
    \includegraphics[width=0.4\textwidth]{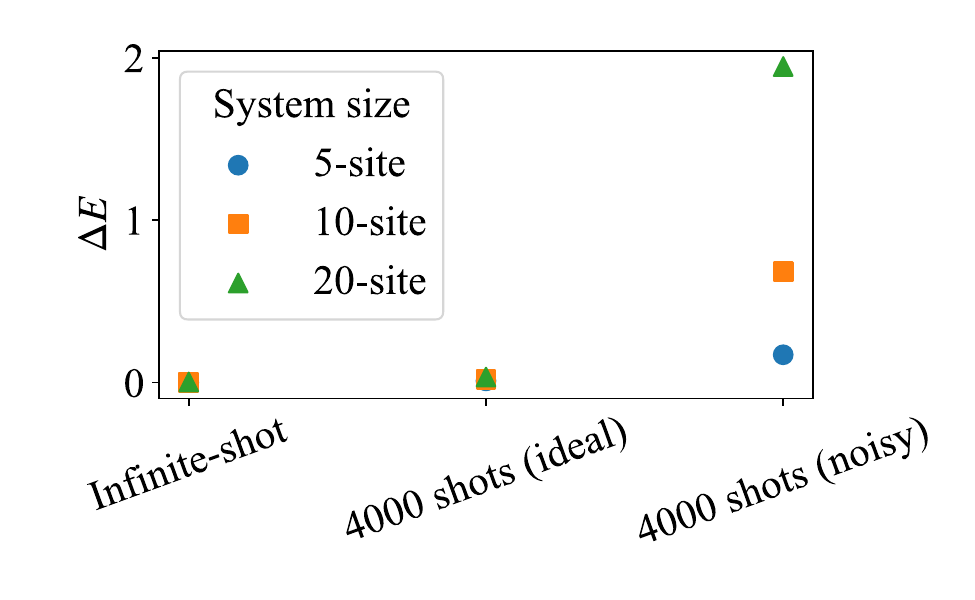}
   \caption{
Energy error $\Delta E = E - E_{\mathrm{exact}}$ for the 5-, 10-, and 20-site
systems at transverse field strength $a = 0.1$, obtained from
infinite-shot and finite-shot (4000) simulations with and without noise.
Different markers indicate system size.
Finite-shot noiseless results remain close to the infinite-shot
reference, while hardware-realistic noise induces a systematic upward
bias that increases with system size.
}
    \label{fig:A4}
\end{figure}

The inclusion of finite-shot noiseless simulations allows the impact of
statistical sampling to be examined independently of hardware noise.
As shown in Fig.~\ref{fig:A4}, across all three system sizes the
finite-shot noiseless results obtained with 4000 measurement shots
remain in close agreement with the infinite-shot reference energies.
In particular, the deviation between infinite-shot and finite-shot
noiseless energies remains below \(0.03\) for all systems considered,
indicating that shot noise alone does not significantly degrade
performance at this sampling level.
In contrast, the introduction of hardware-realistic noise leads to a
systematic upward bias in the estimated ground-state energy, with the
magnitude of the deviation increasing with system size, most notably
for the 20-site ribbon graph.

\clearpage
\bibliographystyle{unsrt}
\bibliography{main}

\end{document}